\documentclass[a4paper,fleqn,usenatbib]{mnras}

\usepackage[T1]{fontenc}
\usepackage{ae,aecompl}

\usepackage{graphicx}	
\usepackage{amsmath}	
\usepackage{amssymb}	
\usepackage[utf8]{inputenc}
\usepackage{ulem}

\renewcommand{\vec}[1]{\mathbf{#1}}
\newcommand{\tensor}[1]{\mathsf{#1}}

\title[Faraday Rotation in the ICM turbulence]{Features of collisionless turbulence in the intracluster medium from simulated Faraday Rotation maps}

\author[M. S. Nakwacki et al.]{
M. S. Nakwacki$^{1,2}$\thanks{E-mail: sole@iafe.uba.ar},
G. Kowal$^{3,4}$,
R. Santos-Lima$^{5}$,
E. M. de Gouveia Dal Pino$^{5}$, \newauthor
and D.~A. Falceta-Gonçalves$^{3,6}$
\\
$^{1}$Instituto de Astronom\'ia y F\'isica del Espacio, UBA-CONICET, Argentina\\
$^{2}$Facultad de Ciencias Exactas y Naturales, Universidad de Buenos Aires, Argentina\\
$^{3}$Escola de Artes, Ci\^encias e Humanidades, Universidade de S\~ao Paulo, S\~ao Paulo, SP, Brazil\\
$^{4}$Núcleo de Astrofísica Teórica, Universidade Cruzeiro do Sul, S\~ao Paulo, SP, Brazil\\
$^{5}$Instituto de Astronom\'ia, Geof\'isica e Ci\^encias Atmosf\'ericas, Universidade de S\~ao Paulo, Brazil\\
$^{6}$SUPA, School of Physics \& Astronomy, University of St Andrews, St Andrews, Fife KY16 9SS, UK
}

\date{Accepted XXX. Received YYY; in original form ZZZ}

\pubyear{2015}

\begin{document}
\label{firstpage}
\pagerange{\pageref{firstpage}--\pageref{lastpage}}
\maketitle

\begin{abstract}
Observations of the intracluster medium (ICM) in galaxy clusters suggest for the
presence of turbulence and the magnetic fields existence has been proved through
observations of Faraday Rotation and synchrotron emission. The ICM is also known
to be filled by a rarefied weakly collisional plasma. In this work we study the
possible signatures left on Faraday Rotation maps by collisionless
instabilities. For this purpose we use a numerical approach to investigate the
dynamics of the turbulence in collisionless plasmas based on an
magnetohydrodynamical (MHD) formalism taking into account different levels of
pressure anisotropy. We consider models covering the sub/super-Alfv\'enic and
trans/supersonic regimes, one of them representing the fiducial conditions
corresponding to the ICM. From the simulated models we compute Faraday Rotation
maps and analyze several statistical indicators in order to characterize the
magnetic field structure and compare the results obtained with the collisionless
model to those obtained using standard collisional MHD framework. We find that
important imprints of the pressure anisotropy prevails in the magnetic field and
also manifest in the associated Faraday Rotation maps which evidence smaller
correlation lengths in the collisionless MHD case. These points are remarkably
noticeable for the case mimicking the conditions prevailing in ICM.
Nevertheless, in this study we have neglected the decrease of pressure
anisotropy due to the feedback of the instabilities that naturally arise in
collisionless plasmas at small scales. This decrease may not affect the
statistical imprint differences described above, but should be examined
elsewhere.
\end{abstract}




\section{Introduction}
\label{sec:introduction}

The intracluster medium (ICM) in galaxy clusters is a very dynamic environment.
Galaxy clusters are built up by gravitational mergers of smaller units according
to the standard scenario of structure formation. They are composed of hundreds
of galaxies in Mpc scale, being the largest viralized structures ($\approx
10^{14}-10^{15}$ M$_\odot$) in the universe. Only $\approx 15 \%$ of the total
mass in galaxy clusters corresponds to baryonic matter, being most of it ($> 80
\%$) in the ICM, and only a small fraction ($< 20 \%$) in stars
\citep{Gonzalez_etal:2007}. The ICM is filled with a hot (at the virial
temperature, $10^{7}-10^{8}$ K) and rarefied gas emitting in the soft-X ray
domain through optically thin bremsstrahlung. Its typical number density ranges
from $0.1$ cm$^{-3}$ to $0.001$ cm$^{-3}$ \citep[][and references
therein]{BanerjeeSharma:2014}.

Modeling of such dynamic environment requires adequate knowledge of the role of
the non-thermal components, namely the relativistic particles and the magnetic
fields. The presence of magnetic fields has been revealed by the study of the
synchrotron emission from diffuse radio sources in the ICM (radio halos and
relics), the Faraday Rotation of the synchrotron emission from radio sources
embedded and behind the cluster \citep[see][]{Ferrari_etal:2008,
Feretti_etal:2012, BrunettiJones:2014}. Through these observations it is
possible to constrain the main properties of the magnetic fields and to
understand the physical processes taking place in the ICM
\citep{Ferrari_etal:2008, Bonafede_etal:2010, Bruggen_etal:2012,
BrunettiJones:2014, Dolag_etal:2005}. For example, radio observations have
discovered the presence of radio emission arising from the ICM and not connected
to the emission of the individual galaxies in the clusters. These radio sources
are called Radio Halos, Radio Relics, and Mini Halos, depending on their
position and observational properties. The emission mechanism is synchrotron
from ultra relativistic electrons diffusing in a turbulent magnetic field at
$\mu$G level. Consequently, a precise knowledge of the ICM magnetic fields could
be an important tool to clarify the origin of the relativistic particles which
are responsible for the synchrotron diffuse radio halos and relics detected in
several galaxy clusters. In particular, observations suggest that radio relics
host relatively large magnetic fields \citep[at $\mu$G
level,][]{Markevitch_etal:2005, vanWeeren_etal:2011}. This magnitude of magnetic
fields is typical in the central regions of clusters, but is not expected near
the periphery, where relics are located.

Several mechanisms have been studied in connection with the amplification of
magnetic fields
\citep{Dolag_etal:2002, Bruggen_etal:2005, Subramanian_etal:2006, IapichinoBruggen:2012}.
A recent work of
\citet{Bruggen:2013} has studied the amplification of magnetic fields due to
shocks including diffusion of cosmic rays, while \citet{SantosLima_etal:2014}
and \cite{Falceta-GoncalvesKowal:2015} have taken into account the pressure
anisotropy of the plasma, thus including the presence of instabilities and
turbulent dynamo to characterize the amplification of the magnetic field. The
impact of such magnetic fields on the Faraday Rotation of the synchrotron
emission is studied in
\citet{BrandenburgStepanov:2014, Sofue_etal:1986, Beck_etal:1996, Fletcher:2010, BeckWielebinski:2013}.
In particular \citet{BrandenburgStepanov:2014} showed that a helical magnetic
field with an appropriate sign of helicity can compensate the so-called Faraday
depolarization, which can have important observational consequences.

Magnetic fields in the ICM also affect the thermal conduction
\cite[see][]{NarayanMedvedev:2001, Lazarian:2006}, given the typical values of
the thermal electron gyro-radius ($\approx 10^8$ cm for $T= 10^8$ K and
$B=1\mu$G) are much smaller than any scale of interest in clusters, and in
particular than the particle mean free path due to collisions \cite[e.g. $\sim
0.05-30$~kpc for Hydra A cluster, 3--7~kpc for Coma cluster, see
e.g.][]{SchekochihinCowley:2006, Andrade-Santos_etal:2013, Sanders_etal:2013}.
It follows that the effective mean free path for diffusion perpendicular to the
magnetic field lines is reduced, and being the magnetic field tangled in the
ICM, it is crucial to obtain information about the magnetic field coherence
length \citep{NakwackiPeraltaRamos:2012, Kunz:2011, SantosLima_etal:2014}.
Magnetic fields in the ICM are also important for the dynamics of cosmic rays,
their diffusion and acceleration \cite[see e.g.][]{Berezinsky_etal:1997,
Petrosian:2001, BrunettiLazarian:2007}.

Usually, the features and phenomena listed above are studied in the context of
standard collisional magnetohydrodynamics. However, the ICM is magnetized and
nearly collisionless, i.e., the gyro-frequency of the ions is greater than that
of binary collision. Plasmas with such characteristics are known to develop
anisotropic pressures with respect to the magnetic field orientation \citep[see
e.g.,][]{QuestShapiro:1996, BarakatSchunk:1982, KrallTrivelpiece:1973}, whose
imprints may survive for considerably long periods compared with the dynamical
timescales of the system \citep{Kowal_etal:2011}. The pressure anisotropy can
give rise to instabilities which are not present in the isotropic case, namely,
firehose and mirror instabilities. In the fluid scales these have a deep impact
on the evolution of turbulence and magnetic field geometry.

In a recent work \citet{SantosLima_etal:2014} have studied the amplification of
magnetic field in the ICM considering the presence of firehose and mirror
instabilities caused by the anisotropic pressure tensor, but including a
plausible model for anisotropy relaxation to mimic the effect of scattering of
individual ions by fluctuations induced by these plasma instabilities in the
kinetic scales. They found that anisotropy in the collisionless fluid is
naturally created by turbulent motions due to fluctuations of magnetic field and
gas density, but the fluctuations of the magnetic field in large scales are
mostly suppressed. The inclusion of such modeling for anisotropy relaxation has
been also considered by \citet{MogaveroSchekochihin:2014}, who found that the
amplification of seed fields is reduced with the use of pressure anisotropy
relaxation.

Improving our theoretical understanding of the dynamics of the ICM plasma can
lead to clearer interpretation of observational data with the aim of
characterizing the magnetic field in the ICM. In particular, numerical
simulations are a useful tool because they allow a control over the free
parameters that are usually considered in observational models. In this respect,
the purpose of this work is twofold. On one hand, we try to give a comprehensive
analysis of numerical simulations of turbulent plasma under different
conditions, including those prevailing in the ICM, taking into account the
effects of pressure anisotropy, and compare the results with those obtained in
the isotropic (MHD) case. On the other hand, we attempt to provide a connection
between the properties of the magnetic field of the ICM that arise using
collisionless and collisional MHD simulations and those usually assumed in
observational studies. In particular, we will consider Faraday Rotation maps and
their statistical properties.

In order to determine the influence of pressure anisotropy on the turbulent
evolution of the magnetic field in the ICM and on the corresponding rotation
maps, in this work we use a collisionless magnetohydrodynamic  formalism with a
double-isothermal closure as implemented in \citet{Kowal_etal:2011}. Considering
that in the ICM the cyclotron frequency is much larger than the collision
frequency, the authors have studied different processes related to pressure
anisotropy. In particular, they used different plasma configurations in order to
determine the appearance of mirror/firehose instabilities and the implications
for the statistics in the velocity field and density \citep{Kowal_etal:2011}.
Here, we do not consider the effects of relaxation of the pressure anisotropy
due to the instabilities feedback upon the plasma as in
\citet{SantosLima_etal:2014}, but will leave this analysis to a forthcoming
work.

This paper is organized as follows. In Section \ref{sec:rm_maps} we give a brief
overview of Faraday Rotation maps and their application to study the magnetic
field structure in the ICM through relevant statistical indicators. In Section
\ref{sec:setup} we describe the theoretical setup, followed by the main results,
presented in Section~\ref{sec:results}, and the Discussions, in
Section~\ref{sec:discussion}. Finally, in Section \ref{sec:conclusions} we
summarize our results and draw the main conclusions.

\section{Faraday Rotation maps of galaxy clusters}
\label{sec:rm_maps}

As mentioned in the previous section, polarization observations of synchrotron
emission have become an important diagnostic tool in the study of the
extragalactic magnetic fields. A brief explanation of this effect is as follows.
When linearly polarized waves, as synchrotron radiation, propagate through a
magneto-ionic medium as the ICM, its polarization properties change. Due to the
birefringence of the medium, in fact, the polarization plane of the radiation is
rotated as a function of frequency. This effect is the so-called Faraday
rotation. If we indicate the intrinsic polarization angle as $\Phi_{int}$, the
effect of the Faraday Rotation can be parametrized as $\Phi_{obs} = \Phi_{int} +
RM\lambda^2$, with the rotation measure (RM) defined as
\begin{equation}\label{eq:RM}
 RM = 812\int_0^Ln_eB_\parallel dl \quad \mathrm{in \, [rad \, m^{-2}]}
\end{equation}
with the magnetic field $B_\parallel$ in $\mu$G, the ambient electronic density
$n_e$ in cm$^{-3}$, both  along the line of sight (LoS), and the distance to the
source $L$ in kpc.

Radio observations of the Faraday Rotation have revealed important features of
the magnetic field, e.g. patchy structures, over a large range of spatial scales
\citep[see e.g.][]{Clarke_etal:2001, CarilliTaylor:2002, Murgia_etal:2004,
GovoniFeretti:2004, VogtEnsslin:2005, Govoni_etal:2010, Bonafede_etal:2010,
Bonafede_etal:2011, KucharEnsslin:2011}. The random magnetic field must be both
tangled on small scales as observed in the RM images and also fluctuate on
scales one or even two orders of magnitude larger, which account for the large
scale turbulence. For this reason, it is necessary to consider cluster magnetic
field models where both small and large scales structures coexist. This behavior
has been studied theoretically on a statistical basis in previous works.
\cite{Murgia_etal:2004} have considered a magnetic fluctuation spectrum in the
form:
\begin{equation}
{|B_{k}|}^2\propto C_n^2 k^{-n}
\label{bk}
\end{equation}
where $n$ represents the spectral index to be constrained by observations and/or
numerical simulations and $C_n^2$ is the power spectrum normalization. The power
spectrum described in Eq.~\ref{bk} has been used in several works proposing a
particular spectral index to reproduce synthetic Faraday Rotation maps, which
usually lies in the range 2--4 \citep[see e.g.][]{Murgia_etal:2004,
Bonafede_etal:2010}, depending on the assumed B--n correlation law.

The origin of these magnetic fields is unknown, but possibly amplified during
the formation of galaxy clusters \citep[e.g.][for reviews]{Dolag:2006,
deGouveiaDalPino:2010, deGouveiaDalPino_etal:2013}. \citet{BhatSubramanian:2013}
considered a fluctuation dynamo action in a turbulent medium to study the
Faraday Rotation measure from the radio emission of background sources seen
through the intermittent magnetic field generated by the dynamo. These authors
showed that even though the magnetic field generated is intermittent, it still
allows the contribution to Faraday Rotation measure to be significant. In this
sense Faraday Rotation measurements are crucial to infer the presence of
coherent magnetic fields and to probe the distribution of the spectral power
over different scales.

In order to track the collisionless effects we investigate the statistical
properties of the Faraday Rotation (FR) maps individually. For this purpose, we
employ statistical tools like the probability density function (PDF) and the
power spectrum of the FR maps, aiming at analyzing these maps for different
regimes of turbulence, always comparing with the output from collisional MHD
models. We will also present the autocorrelation function of the synthetic FR
maps, which is more directly connected with observations.

\section{Theoretical setup and numerical simulations}
\label{sec:setup}

\subsection{Double-isothermal collisionless MHD approximation}
\label{sec:setup:equations}

The double-isothermal collisionless MHD approximation consists of a single fluid
plasma with the pressure replaced by the pressure tensor described by two
isothermal equation of states with independent isothermal sound speeds along the
parallel and perpendicular direction to the local magnetic field. The
conservation laws which describe this approximation
\cite[e.g.,][]{Kowal_etal:2011}\footnote{We note that there are other approaches
to collisionless MHD, of which the so called Chew-Golberger-Low closure (CGL)
\citep{Chew_etal:1956} is the most employed. The main difference between the CGL
and the present double-isothermal approach is that the first combines the
adiabatic conservation of magnetic momentum of the particles with local thermal
energy conservation (that is, the conservation of the local entropy) while the
second keeps constant temperatures of the gyrotropic distribution of particle
velocities in parallel and perpendicular directions to the local magnetic field
\cite[see][]{Kowal_etal:2011}.} are:
\begin{eqnarray}
 \frac{\partial\rho}{\partial t} + \nabla \cdot \left( \rho\vec{v} \right) & = & 0, \label{eq:mass} \\
 \frac{\partial\left(\rho\vec{v}\right)}{\partial t} + \nabla \cdot \left[ \rho\vec{v}\vec{v} + \left( \tensor{P} + \frac{B^2}{8\pi} \right) \tensor{I} - \frac{\vec{B}\vec{B}}{4\pi} \right] & = &\vec{f}, \label{eq:momentum} \\
 \frac{\partial\vec{B}}{\partial t} - \nabla \times \left( \vec{v} \times \vec{B} \right) & = & 0 \label{eq:induction},
\end{eqnarray}
where the pressure tensor is described by two components:
\begin{eqnarray}\label{eq:pressure}
 \tensor{P} = p_\perp \tensor{I} + (p_\parallel - p_\perp)\hat{b}\hat{b}, \quad
 p_{\perp} = c^2_{\perp} \rho \quad {\rm and} \quad
 p_{\parallel} = c^2_{\parallel} \rho,
\end{eqnarray}
with $\hat{b} = \vec{B}/|\vec{B}|$, and $c_{\parallel, \perp}$ being the sound
speeds parallel and perpendicular to the magnetic field $\vec{B}$, respectively.
Then, the momentum conservation equation can be written as:
\begin{equation}\label{Eq.2}
 \frac{\partial(\rho\vec{v})}{\partial t} + \nabla \cdot \left[ \rho \vec{v} \vec{v} + \left( c_\perp^2\rho + \frac{B^2}{8\pi} \right) \tensor{I} - \left( 1 - \alpha \right) \vec{B}\vec{B} \right] = \vec{f},
\end{equation}
where $\alpha = \frac{1}{2}(p_\parallel-p_\perp) / p_m$, and $p_m= {B^2}/{8\pi}$
is the magnetic pressure.

A linear analysis \citep[see][]{HauWang:2007, Kowal_etal:2011} provides the
following dispersion relation for the waves (of wavenumber $k$ and frequency
$\omega$):
\begin{eqnarray}\label{eq:lin_waves}
\left( \frac{\omega}{k} \right)^2_{A} = \left[ V^2_A - (c^2_{\parallel} - c^2_{\perp}) \right] \cos^2 \theta, \\
\left( \frac{\omega}{k} \right)^2_{F,S} = \frac{1}{2} \left( c^2_{\perp} + V^2_{A} \pm \sqrt{\Delta} \right),
\end{eqnarray}
where
\begin{eqnarray}\nonumber
  \Delta = \left( c^2_{\perp} + V^2_{A} \right)^2
- 4 \left[ c^2_{\perp} \left( c^2_{\parallel}- c^2_{\perp} \right)+ c^2_{\parallel} V^2_{A} \right] \cos^2 \theta \\
+ 4 \left( c^4_{\parallel} - c^4_{\perp} \right) \cos^4 \theta,
\label{eq:delta}
\end{eqnarray}
and the subscripts $A, F, S$ stand for the Alfv\'en, fast, and slow modes,
respectively, in analogy to the standard MHD case. It can be shown that $\Delta
> 0$ always. The usual MHD dispersion relations are recovered for $c_{\perp} =
c_{\parallel}$.

The first thing to note in the equations above is that now the linear Alfv\'en
wave depends also on the thermal speeds. Also the maximum and minimum values of
the fast mode change and can be larger than in the case of isotropic pressure.
As the function $\Delta$ is not anymore linear in $\cos^2 \theta$, the maximum
and minimum speeds of the fast mode can be now between the extremes $\theta = 0$
or $\theta = \pi$. The same applies to the maximum speed of the slow modes.
Another difference introduced by the anisotropy is the possibility of anomalous
slow modes with $\delta b_{\parallel} \delta \rho > 0$, that is, with a positive
correlation between the density fluctuations and the magnetic field component
parallel to the background magnetic field (see \citet{HauWang:2007} and
Appendix~\ref{appendix:dispersion}).

The most remarkable difference introduced by the anisotropy is the possibility
of occurrence of instabilities. When $c_{\parallel}/c_{\perp}>1 $ the firehose
instability can arise which tends to bend the magnetic field lines and trap gas
within regions of high intensity field. On the other hand, when
$c_{\parallel}/c_{\perp}<1 $ the mirror instability can occur pushing gas to
regions of smaller magnetic field strength \citep[e.g.][]{Kulsrud:1983}. A brief
description of the conditions for each instability is given in
Appendix~\ref{appendix:instabilities}.

\subsection{The numerical code}
\label{sec:setup:code}

In order to study the magnetic field dynamics in the ICM we simulate turbulence
in a three-dimensional periodic Cartesian box evolving the double-isothermal
collisionless MHD equations (\ref{eq:mass}--\ref{eq:pressure}). The turbulence
is introduced through a source term $\vec{f}$ on the right-hand side of
Eqs.~(\ref{eq:momentum}) and (\ref{Eq.2}) and is continuously driven at the wave
scale $k= 2.5$ (which gives an injection scale in the model 2.5 times smaller
than the size of the computational box). Our forcing is done in Fourier space in
such a way, that the forcing components have random phases at each step and the
correlation between them and velocity is removed. This assures that the velocity
field is not correlated with our forcing at any temporal or spatial scale.
Additionally, our forcing is incompressible, and therefore does not generate
density fluctuations by itself. Any compression appearing in the studied systems
is the result of the magnetosonic wave interactions or kinetic instabilities
developed during the system evolution \cite[see e.g.][and references
therein]{Kowal_etal:2007, KowalLazarian:2010}.  As turbulence evolves, the MHD
modes interact and generate both compressive and incompressive components at
scales smaller than the injection one. Therefore, turbulence observed in our
models is actually mixture of both modes.

To solve Eqs. (\ref{eq:mass}--\ref{eq:pressure}) we employ the shock-capturing,
second order Godunov code \cite[see][for more detailed code
description]{Kowal_etal:2007, Kowal_etal:2009, KowalLazarian:2010,
Kowal_etal:2011}. We do not take into account viscosity or diffusion in the
equations. The scale at which the dissipation starts to act is defined by the
numerical diffusivity of the scheme (see
Section~\ref{sec:discussion:turbulence}).

The variables are normalized in code units therefore they can be rescaled to any
system by defining three representative quantities from which all the other ones
can be derived: the length scale $L$ (which is given by the computational box
size), the gas density $\rho_0$ (given by the initial ambient density of the
system), and the Alfv\'en speed defining the time unit $L/V_A$. With this
normalization, the isothermal sound speeds ($c_s$, $c_\parallel$, and $c_\perp$)
are also given in units of $V_A$ and the magnetic field has units $V_A
\sqrt{4\pi \rho_0}$.

\subsection{Initial conditions}
\label{sec:setup:conditions}

In order to understand better the connection between some features of the
magnetic field and the FR maps, it is useful to consider different initial
conditions for the simulations, corresponding to different regimes in which the
turbulent plasma evolves. Specifically, we shall analyse the impact of pressure
anisotropy on the maps by considering different initial conditions for the
magnetic field strength, and parallel/perpendicular pressure anisotropy.

We consider six different models that cover the sub/super-Alfv\'enic and the
trans/supersonic regimes; the respective parameters are shown in Table
\ref{tab:models}. The turbulent velocity at the injection scale is $V_{turb}
\approx 0.8$. It must be remarked that the classifications of
sub/super-Alfv\'enic and trans/supersonic refer to the collisional MHD models
(which are employed for comparison with the collisionless models). These are
based on the comparison between $V_{turb}$ and the Alfv\'en speed $V_A =
B_{ext}/\sqrt{\rho_0}$ (in code units), and $V_{turb}$ and the parallel sound
speed $c_{\parallel}$, respectively.  Indeed, as discussed in
Section~\ref{sec:setup:equations}, an anisotropic ``super-Alfv\'enic''
($V_{turb} > V_A$) model can be effectively sub-Alfv\'enic depending on the
sound speed. The comparison of the results obtained from these models allows us
to get a deeper insight into the effect of the magnetic field topology on the
Faraday Rotation maps. We consider the initial magnetic field in the $\hat{x}$
direction, and an initial constant density $\rho_0$ set to $1.0$ in code units
for all six models. We evolve up to $t=5.0$ in code units, when the turbulence is
fully developed (we note that the turbulence turn-over time is $\sim 0.4$ in
code units).

Figure~\ref{fig:dispersion} in Appendix~\ref{appendix:dispersion} shows the
linear phase speeds (eq.~\ref{eq:lin_waves}) for the initial condition of each
of the models in Table \ref{tab:models}. It compares the wave speeds of the
models with anisotropic pressure with those of isotropic (collisional) MHD
models. It highlights, for example, the differences in the effective Alfv\'en
speeds (which also reflect the magnetic field tension).
Figure~\ref{fig:dispersion} also reveals what models are initially unstable to
the firehose and mirror instabilities. Table~\ref{tab:instabilities} shows the
magnetic field intensity which is required for each model to develop firehose or
mirror unstable regions (see instability conditions in
Appendix~\ref{appendix:instabilities}). It also classifies the turbulence as
super- or sub-Alfv\'enic and trans- or subsonic according to the corresponding
collisional MHD model that is used for comparison with each collisionless model.
The last column shows the effective Alfv\'en speed, that is, the phase speed of
the linear Alfv\'en wave for the initial conditions.

As indicated in Table \ref{tab:models}, six collisionless MHD models are
considered. Models 1 and 2 are in the transonic and sub-Alfv\'enic regime. Model
1 is initially unstable to mirror modes (see Table \ref{tab:instabilities} and
Figure~\ref{fig:dispersion}). Model 2 is initially stable to both mirror and
firehose instabilities, but later on, in locations where the magnetic field is
reduced to values smaller than the threshold (Table \ref{tab:instabilities}), it
can develop the firehose instability. Figure~\ref{fig:dispersion} shows that the
effective behaviour of model 2 is super-Alfv\'enic as the initial effective
Alfv\'en speed is reduced with respect to $V_A$. Models 3 and 4 correspond to
supersonic and super-Alfv\'enic regimes. Model 3 is also unstable to mirror
modes, and Model 4 can also later develop flow regions which are unstable to the
firehose instabilities, if the magnetic field intensity is reduced to values
smaller than the threshold for its triggering (Table \ref{tab:instabilities}).
Model 5 is initially in a transonic and super-Alfv\'enic turbulence regime and
can eventually develop firehose instabilities in regions where the magnetic
field is reduced to values below the threshold for this model (see Table
\ref{tab:instabilities}). This model may represent the physical regime
prevailing in compressed zones of the ICM.\footnote{We note that
\cite{SantosLima_etal:2014} examined the conditions of the turbulent
collisionless plasma in the ICM and found that the flow tends to develop regions
predominantly with $c_{\perp} > c_{\parallel}$ (which favours the triggering of
mirror instabilities) and compressed regions with $c_{\parallel} > c_{\perp}$
(which favours the onset of the firehose instability).} Finally, Model 6 is
initially in the supersonic and sub-Alfv\'enic turbulent regime and is stable to
both mirror and firehose instabilities. We also note that since this model is
strongly sub-Alfv\'enic, it is very unlikely that later on it will be able to
produce magnetic fields as small as those required for the onset of the mirror
instability (see Table \ref{tab:instabilities}). Besides, the plasma
$\beta=p/p_m$ ratio is very low in this case so that the pressure anisotropy
will have little dynamical importance on the flow.

The collisional MHD models which are considered for comparison with the
collisionless counterparts are also listed in Table~\ref{tab:models}. They have
four different combinations between initial magnetic field strength and sound
speed in order fulfil the same initial conditions of the models mentioned above.

We should note that in the ICM the density and temperature profiles are not
constant and decay with the distance from the cluster core. Following the model
of gas distribution in the relaxed ICM by \cite{CavaliereFuscoFemiano:1978}, the
density can drop by factor 10--100 at distance of 5 core radii. Similar models
are being studied for the pressure profiles providing comparable decay
\cite[see][]{Lapi_etal:2012}. As the result, the sound speed should not change
by more than a factor 2--3 at the distance of 1--2~Mpc from the core. While the
profiles for density and temperature can be obtained from the X-ray
observations, the profile for large scale component of magnetic field is poorly
know. The problem comes from the fact that in order to obtain the magnetic field
strength from RM observations, one has to know the profile of density, which is
obtained from the surface density using the mentioned models.

Our numerical models represent local simulation of a box embedded in the ICM far
from local sources with its size corresponding to a fraction of the cluster
sizes, of the order of tens of kpcs compared to the size of clusters of
1--2~Mpc. It means, that the change of global profiles is small within our
numerical domain. In addition, we assume that turbulence is driven at scales
larger than our simulation box. These two assumptions justify the use of
periodic boundary conditions. We are planning, however, to extend our studies in
the future by introducing external profiles for density, pressure, and magnetic
field, which would also require inclusion of the gravitational potential of the
cluster and better suited boundary conditions, such as hydrostatic ones.

The physical characteristics of all models are discussed below where we present
our results in connection with the imprints left on Faraday Rotation maps
(Section \ref{sec:results}).

\begin{table}
\centering
\begin{tabular}{crrrr}
\hline\hline
Model & $B_{ext}$ & $c_\parallel$ & $c_\perp$ & $c_\parallel^2/c_\perp^2$ \\
\hline\hline
\multicolumn{5}{c}{collisional models} \\
1-2 & 1.0 & 1.0 & 1.0  & 1.0 \\
3-4 & 0.1 & 0.1 & 0.1  & 1.0 \\
5   & 0.1 & 1.0 & 1.0  & 1.0 \\
6   & 1.0 & 0.1 & 0.1  & 1.0 \\
\hline
\multicolumn{5}{c}{collisionless models} \\
1   & 1.0 & 1.0 & 2.0  & 0.25 \\
2   & 1.0 & 1.0 & 0.5  & 4.0  \\
3   & 0.1 & 0.1 & 0.2  & 0.25 \\
4   & 0.1 & 0.1 & 0.05 & 4.0  \\
5   & 0.1 & 1.0 & 0.5  & 4.0  \\
6   & 1.0 & 0.1 & 0.2  & 0.25 \\
\hline
\end{tabular}
\caption{Parameters of the turbulent isothermal collisional and
double-isothermal collisionless MHD models simulated with the resolution
$512^3$. The initial density for all models was set to 1.0. For collisional
models, the parallel and perpendicular sound speeds are equal, i.e. $c_\parallel
= c_\perp = c_{snd}$. Their names indicate the corresponding collisionless
models for which both $B_{ext}$ and $c_\parallel$ are the same.}
\label{tab:models}
\end{table}

\begin{table*}
\centering
\begin{tabular}{crccccc}
\hline\hline
Model & $B_{ext}$ & mirror & firehose & $V_{A0,eff}$ & sonic regime & Alfv\'enic regime \\
\hline\hline
1 & 1.0 & $B<3.4$  & --        & 2.0  & trans & sub   \\
2 & 1.0 & --       & $B<0.86$  & 0.5  & trans & sub   \\
3 & 0.1 & $B<0.34$ & --        & 0.2  & super & sub   \\
4 & 0.1 & --       & $B<0.086$ & 0.05 & super & super \\
5 & 0.1 & --       & $B<0.86$  & --   & trans & super \\
6 & 1.0 & $B<0.34$ & --        & 1.0  & super & sub   \\
\hline
\end{tabular}
\caption{Characteristics of the simulated models from Table~\ref{tab:models}. In
the first and second column we show the model name and its mean magnetic field
strength. The third and fourth columns show the local magnetic field intensity
required to develop the mirror or firehose instabilities, respectively. The next
column shows the effective Alfv\'en speed for the initial conditions of the
collisionless models ($V_{A0,eff}^2 = V_{A0}^2 + c^2_{\perp} -
c^2_{\parallel}$). The last two columns show the sonic and Alfv\'enic regimes of
the developed turbulence.}
\label{tab:instabilities}
\end{table*}

\section{Results}
\label{sec:results}

In this section we present the results obtained from the simulations with
parameters given in Table~\ref{tab:models} (Models 1--6).  In order to extract
the main features that the anisotropy imprints on the plasma, we analyze the
magnetic field intensity, its probability distribution, and energy power
spectrum.  We present the synthetic Faraday Rotation maps, and perform a
detailed statistical analysis focusing on the characteristics of their
distributions and power spectra.  To complete our analysis, we also present the
autocorrelation functions for these maps, which can be compared directly with
observations.

Through this section, all presented results consider only the last snapshot
corresponding to the time of the simulation $t=5.0$, when the turbulent cascade
has been already fully developed.

\subsection{Magnetic field intensity and the role of anisotropy}
\label{sec:results:field_intensity}

\begin{figure*}
\centering
\includegraphics[width=\textwidth]{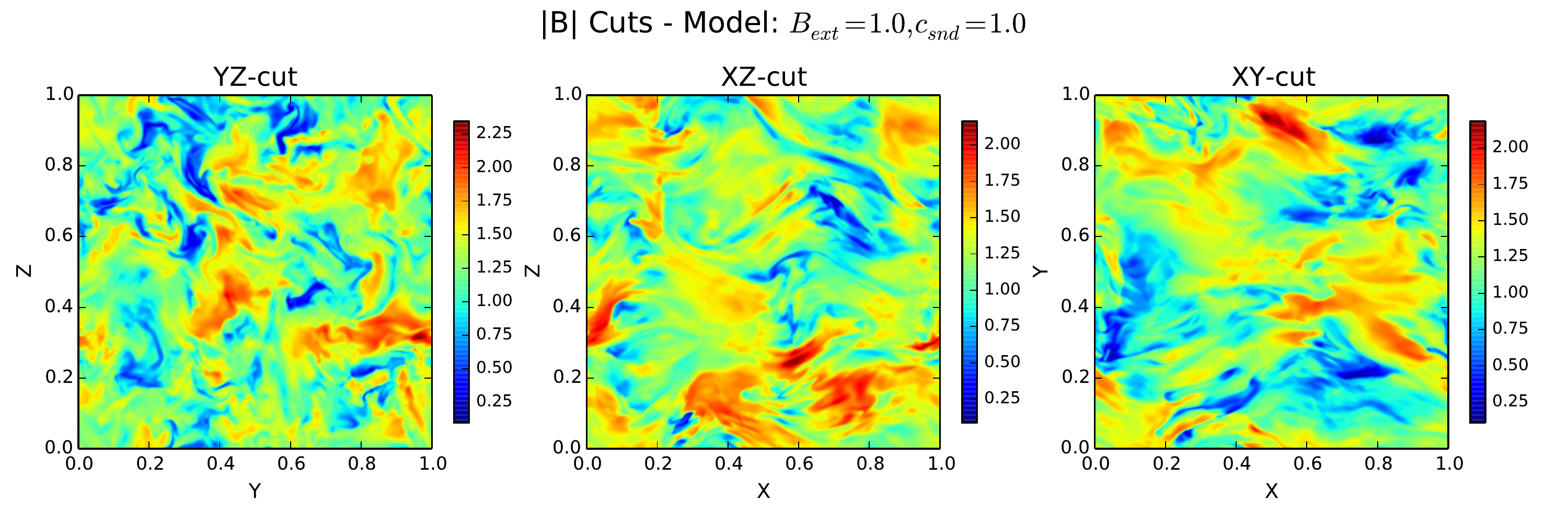}
\includegraphics[width=\textwidth]{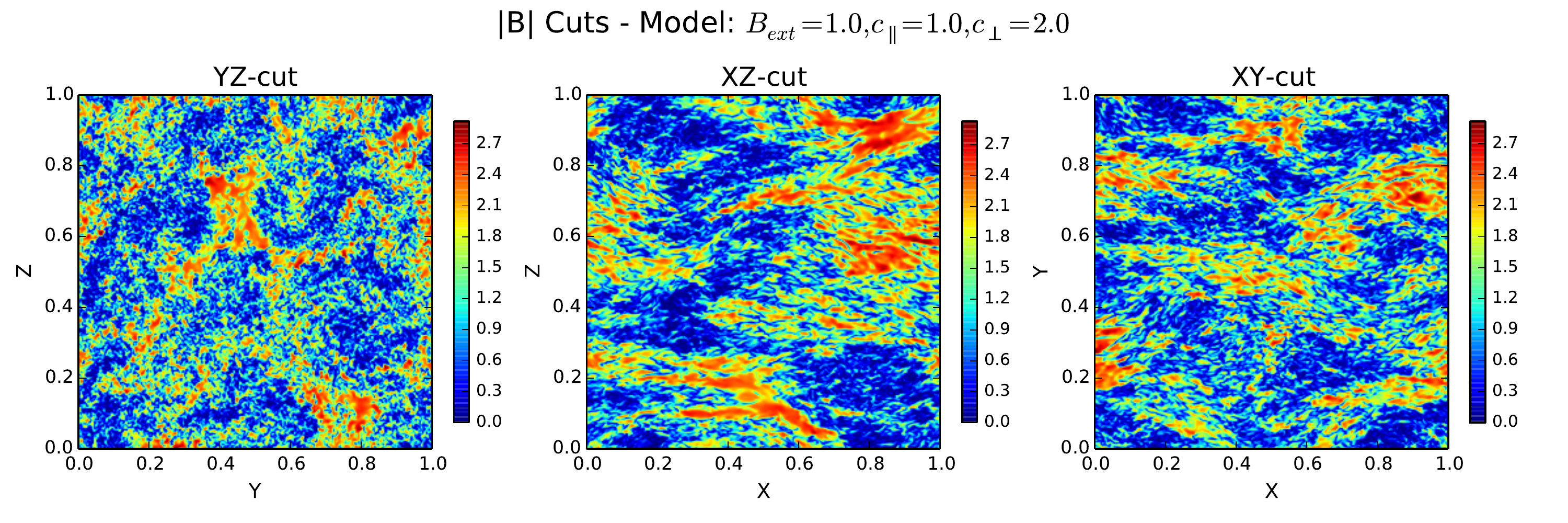}
\includegraphics[width=\textwidth]{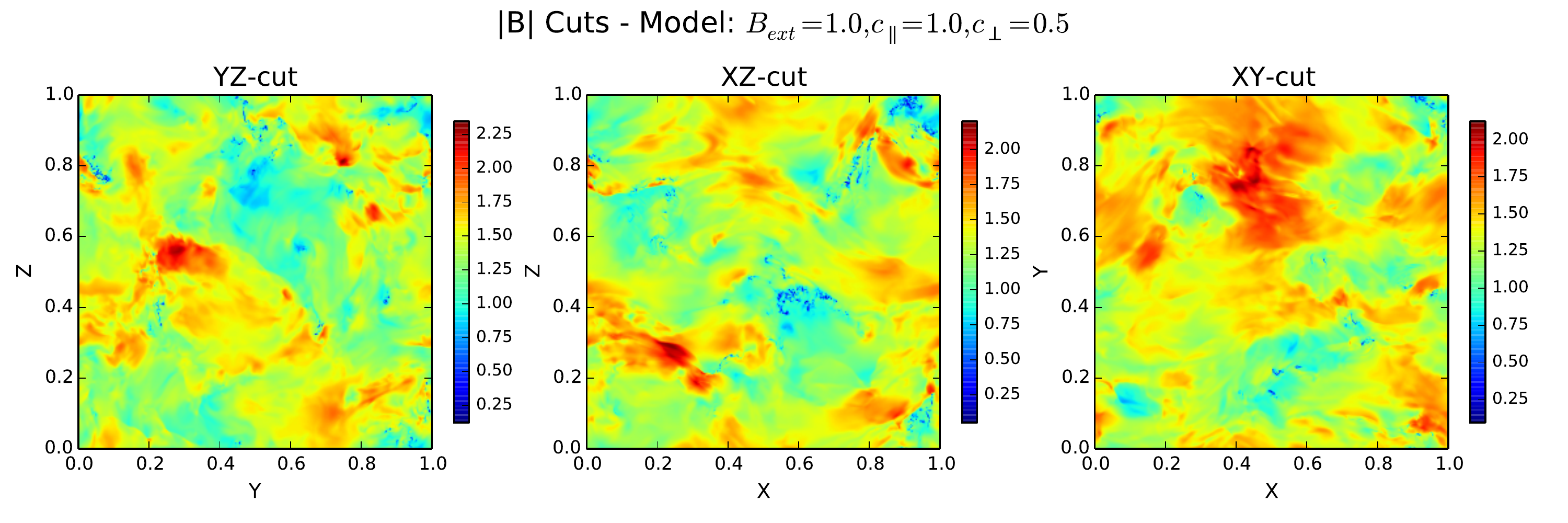}
\caption{Central slices across the computational domain showing the magnetic
field intensity $|\vec{B}|$ with normal direction along $\hat{x}$, $\hat{y}$,
and $\hat{z}$ (left, middle, and right panels, respectively) for collisional
Model 1-2 (upper row), and collisionless Models 1 and 2 (middle and lower rows,
respectively). Each plot has its individual color scale.}
\label{f1b1}
\end{figure*}

\begin{figure*}
\centering
\includegraphics[width=\textwidth]{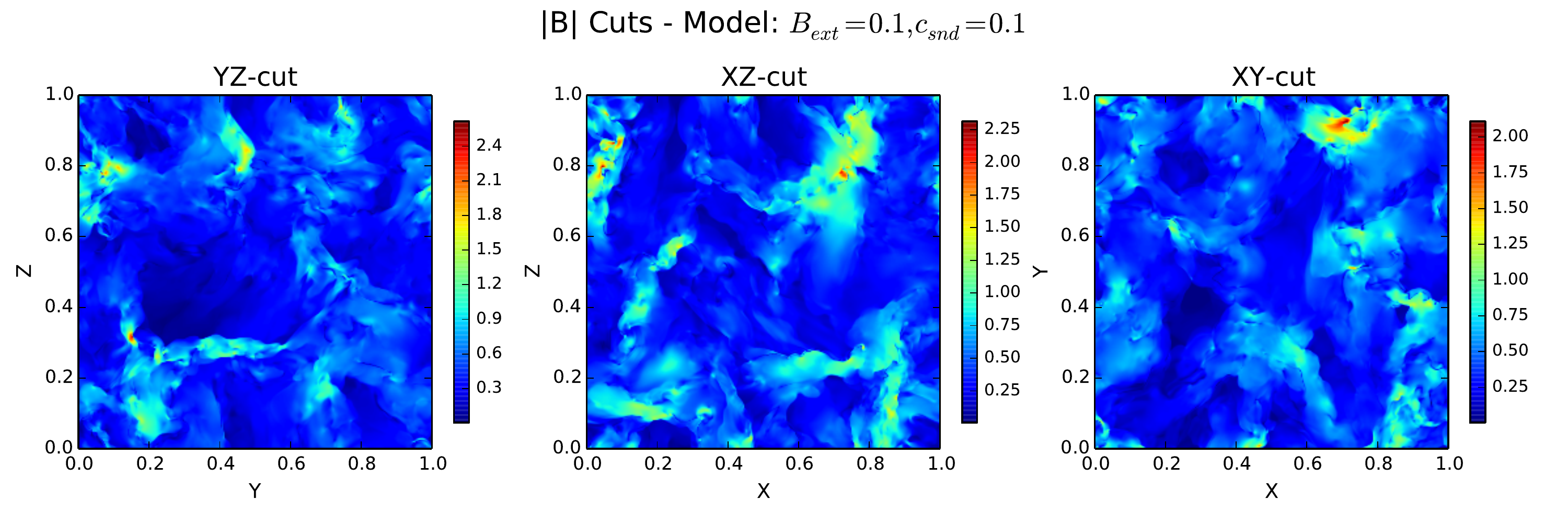}
\includegraphics[width=\textwidth]{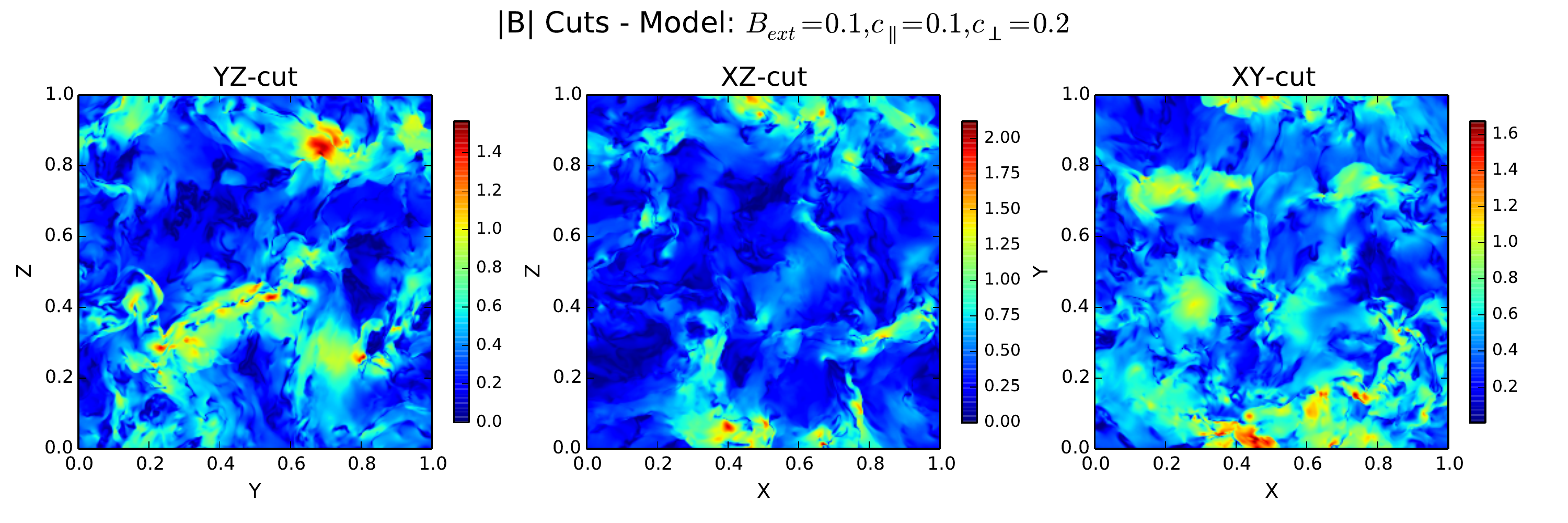}
\includegraphics[width=\textwidth]{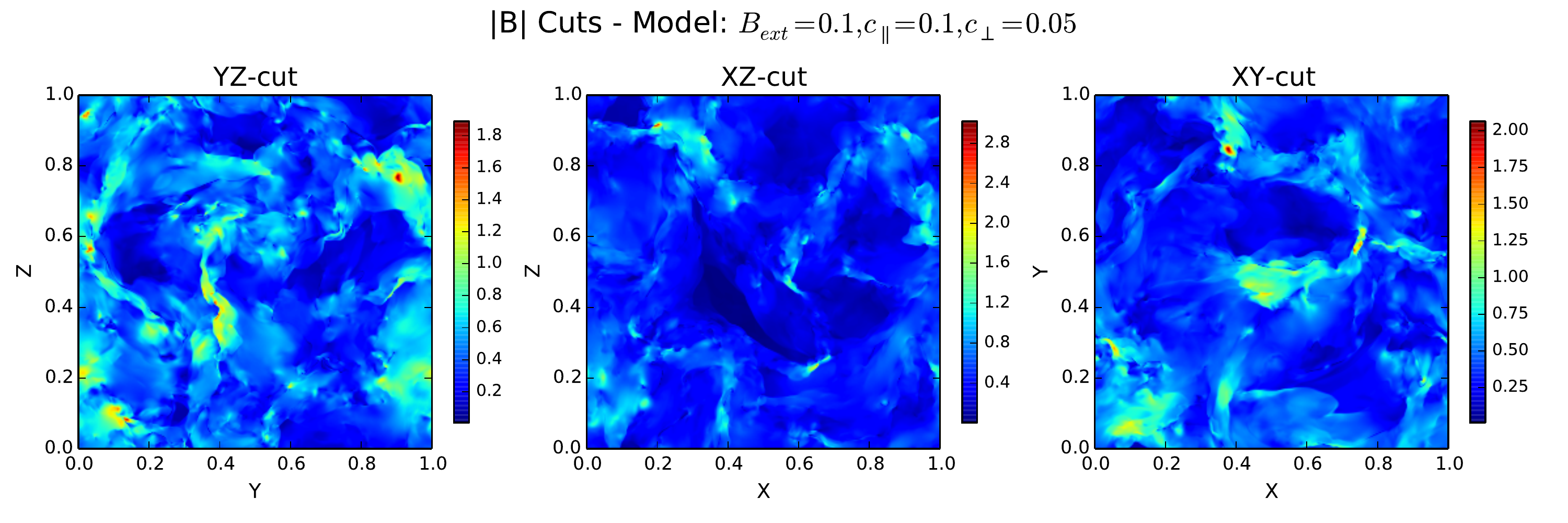}
\caption{Central slices across the computational domain showing the magnetic
field intensity $|\vec{B}|$ for collisional Model 3-4 (upper row), and
collisionless Models 3 and 4 (middle and lower rows, respectively).}
\label{f1b2}
\end{figure*}

\begin{figure*}
\centering
\includegraphics[width=\textwidth]{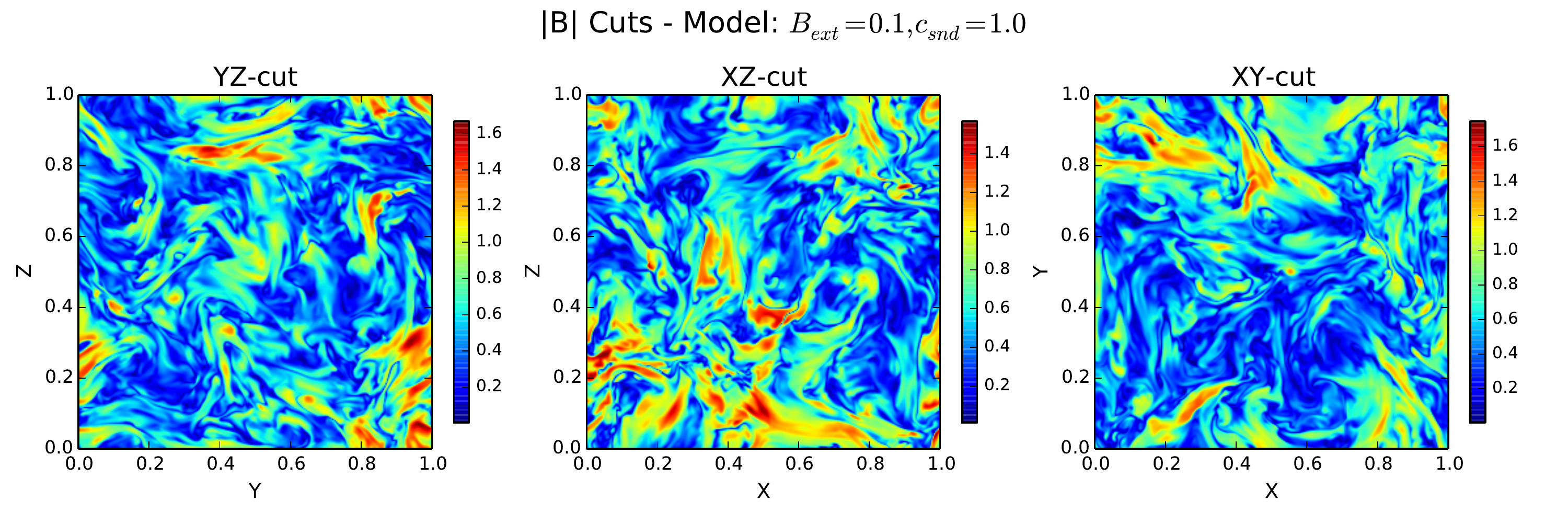}
\includegraphics[width=\textwidth]{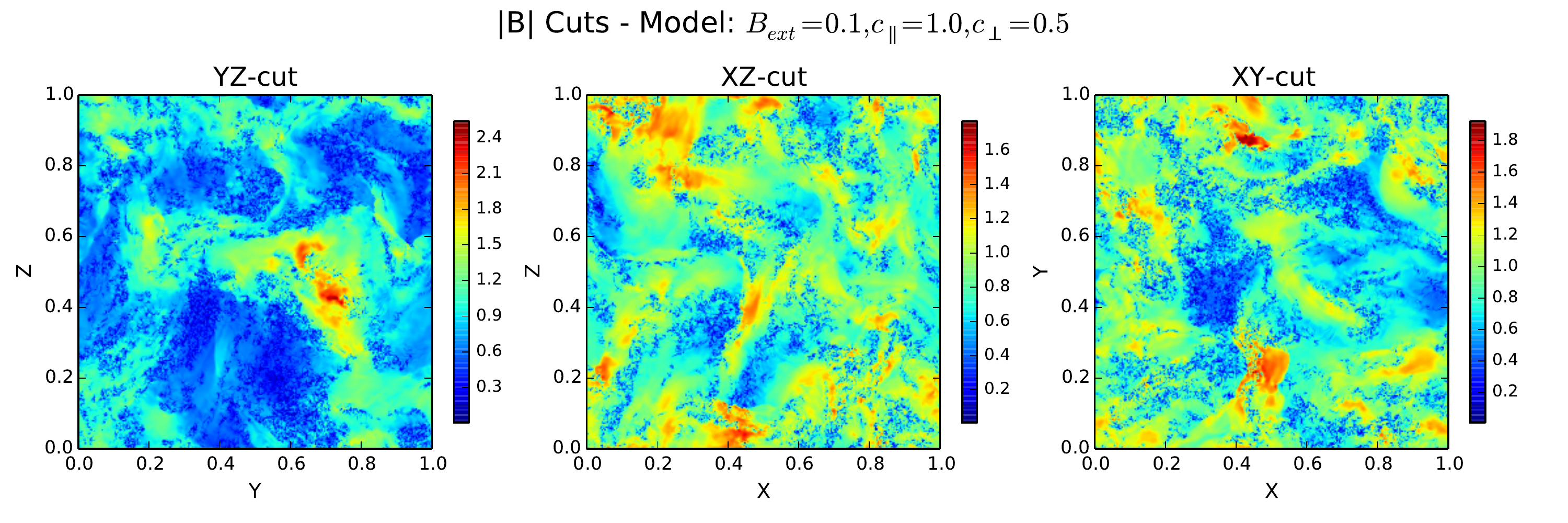}
\caption{Central slices across the computational domain showing the magnetic
field intensity $|\vec{B}|$ for collisional and collisionless Model 5 (upper
and lower rows, respectively).}
\label{f1b3}
\end{figure*}

\begin{figure*}
\centering
\includegraphics[width=\textwidth]{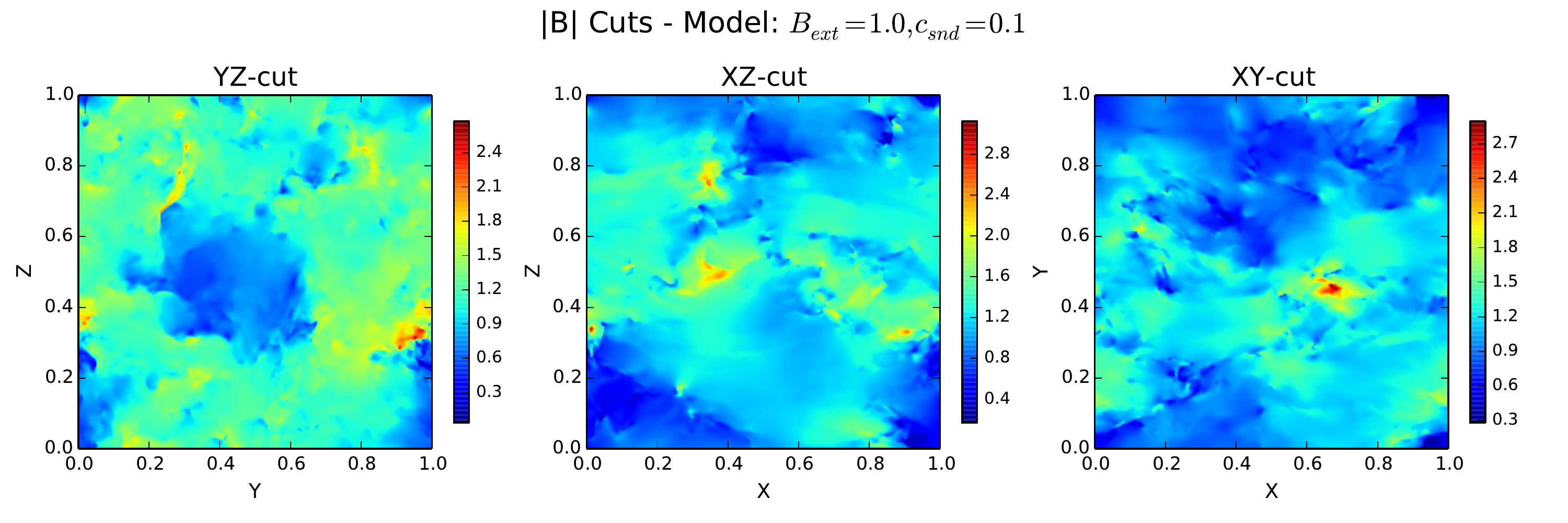}
\includegraphics[width=\textwidth]{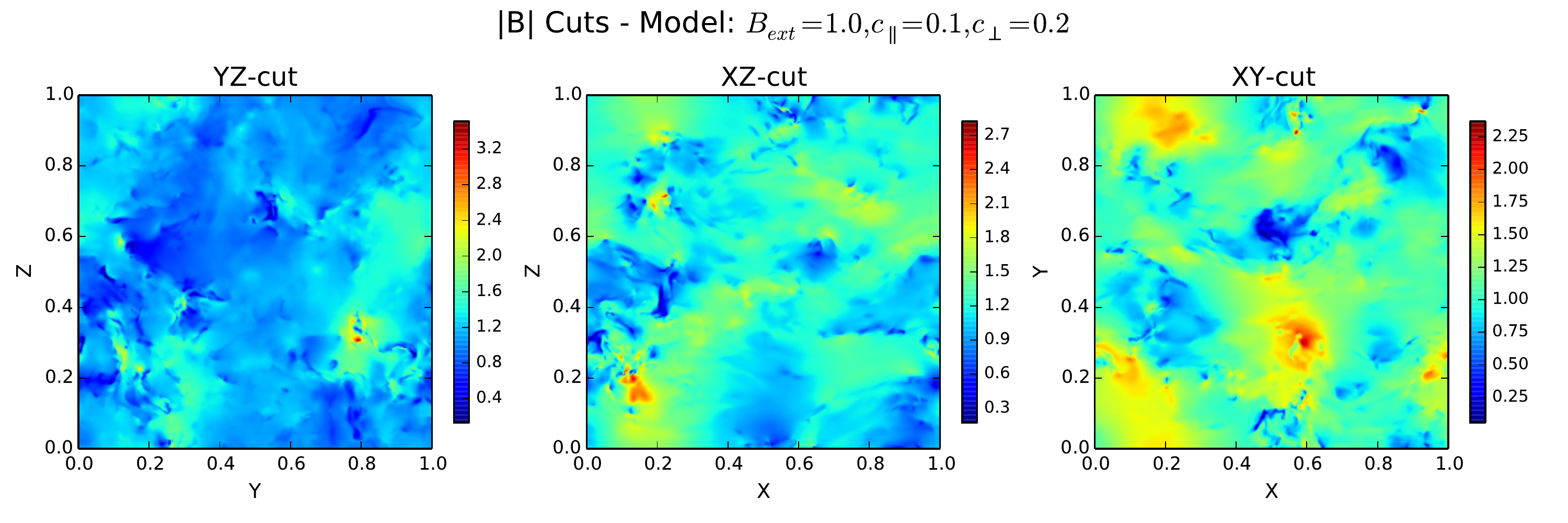}
\caption{Central slices across the computational domain showing the magnetic
field intensity $|\vec{B}|$ for collisional and collisionless Model 6 (upper
and lower rows, respectively).}
\label{f1b4}
\end{figure*}

Figures~\ref{f1b1}--\ref{f1b4} depict a cut of the magnetic field intensity in
the center of the computational domain for all six models from
Table~\ref{tab:models}. In these figures we compare the MHD models with
isotropic (top rows) and anisotropic pressure (middle and/or bottom rows).

In the case of the transonic and sub-Alfv\'enic regime, corresponding to Model 1
(second row of Fig.~\ref{f1b1}), the presence of mirror instability produces
strong fluctuations of magnetic field in small scales, which can grow since the
turbulent motions are slow compared to the instability growth rate (see
Fig~\ref{fig:dispersion}). The instability presence is seen in large differences
between the magnetic field structure obtained in Model~1 and the corresponding
model with the isotropic pressure (seen in the top row of Figure~\ref{f1b1}). As
pointed out in \citet{Kowal_etal:2011}, due to the instability growth rate
larger at small scales, the instability creates more granulated maps, thus the
magnetic field intensity for Model~1 results in a more wrinkled distribution
\footnote{In our numerical simulations, the numerical dissipation suppresses the
instabilities in the dissipation range. See
Section~\ref{sec:discussion:turbulence}.}. The more intense magnetic field
regions are distributed in smaller patches. Moreover, the fragmented structures
seem to be more aligned with the initial magnetic field.

The transonic and sub-Alfv\'enic turbulent Model 2 is shown in the bottom row of
Figure \ref{f1b1}. Even though this model has an anisotropic stress smaller than
the magnetic pressure (i.e., $\lvert p_{\parallel} - p_{\perp} \rvert / 2 p_m <1
$) it is enough to reduce the magnetic tension and make the turbulence
effectively super-Alfv\'enic (see the effective Alfv\'en speed in
Table~\ref{tab:instabilities}). Only in regions where the magnetic field
intensity is reduced to values below the threshold
(Table~\ref{tab:instabilities}) the plasma can become unstable, producing the
small-scale structures observed in some parts of the map.  In general, this
model is similar to the collisional MHD case.

The cases corresponding to supersonic and super-Alfv\'enic regimes are shown in
Figure \ref{f1b2}, and the turbulence in these cases is dynamically dominant
over the growth of the instabilities. The differences in the magnetic field
intensity distribution between the collisional (upper row of Fig. \ref{f1b2})
and collisionless MHD models (middle and bottom rows of Fig. \ref{f1b2}) are
small, being more noticeable by the presence of more structures (at small
scales) in Model 3, which becomes mirror unstable for $B<0.34$ (see
Table~\ref{tab:instabilities}). Model 4, however, is practically stable over its
whole volume (the firehose modes can arise only for $B<0.086$ in this case). It
can be also observed that in Model 3 the intensities of the magnetic field are
smaller than in the corresponding case with isotropic pressure. This is due to
the larger effective Alfv\'en speed, which reduces the magnetic field
amplification due to turbulence.

Figure \ref{f1b3} shows the magnetic field intensity corresponding to the
transonic and super-Alfv\'enic regime. Model 5 can represent qualitatively the
conditions in compressed zones in the ICM. The strong firehose instability
deforms the magnetic field lines, decreases the anisotropy of fluctuations with
respect to the magnetic field lines, and produces a very granulated distribution
of the magnetic field intensity that is very different from the structure found
in the collisional MHD model. The curved magnetic lines tend to slow down and
trap the flowing gas in regions of larger magnetic field.\footnote{We note that
the LoS along the initial magnetic field can be easily distinguished from the
LoSs along other directions (see the right column). This is, however, just an
effect of the different color scaling. As a matter of fact, in the presence of
the firehose instability, the turbulence becomes more isotropic because the
field lines cannot resist to bending as in super-Alfv\'enic turbulence.} As in
Model 1, the firehose instability can freely grow without being suppressed by
the turbulent motions of the gas. It is responsible for the generation of
small-scale magnetic field fluctuations and tangling the field lines, which
result in an increase of the perpendicular pressure in the local reference
frame. Comparing the cases with isotropic and anisotropic pressures, it is seen
that in the former the magnetic field intensity is more elongated, and in the
latter, several regions of larger magnetic field are formed, while the less
intense magnetic field zones are confined to smaller regions.

The magnetic field distribution obtained from Model 6 (corresponding to a
supersonic and sub-Alfv\'enic turbulent regime) is shown in Fig. \ref{f1b4}. In
this case the collisionless and collisional MHD models are very similar because
the sub-Alfv\'enic turbulence is unable to drive the plasma in Model~6 into the
mirror unstable regime given by the condition $B<0.34$ (see
Table~\ref{tab:instabilities}), in which case the thermal pressure anisotropy is
dynamically unimportant and does not disturb the turbulent motions.

\subsection{Probability distribution function and power spectra of the magnetic field}
\label{sec:results:field_pdfs_spectra}

\begin{figure*}
\centering
\includegraphics[width=0.48\textwidth]{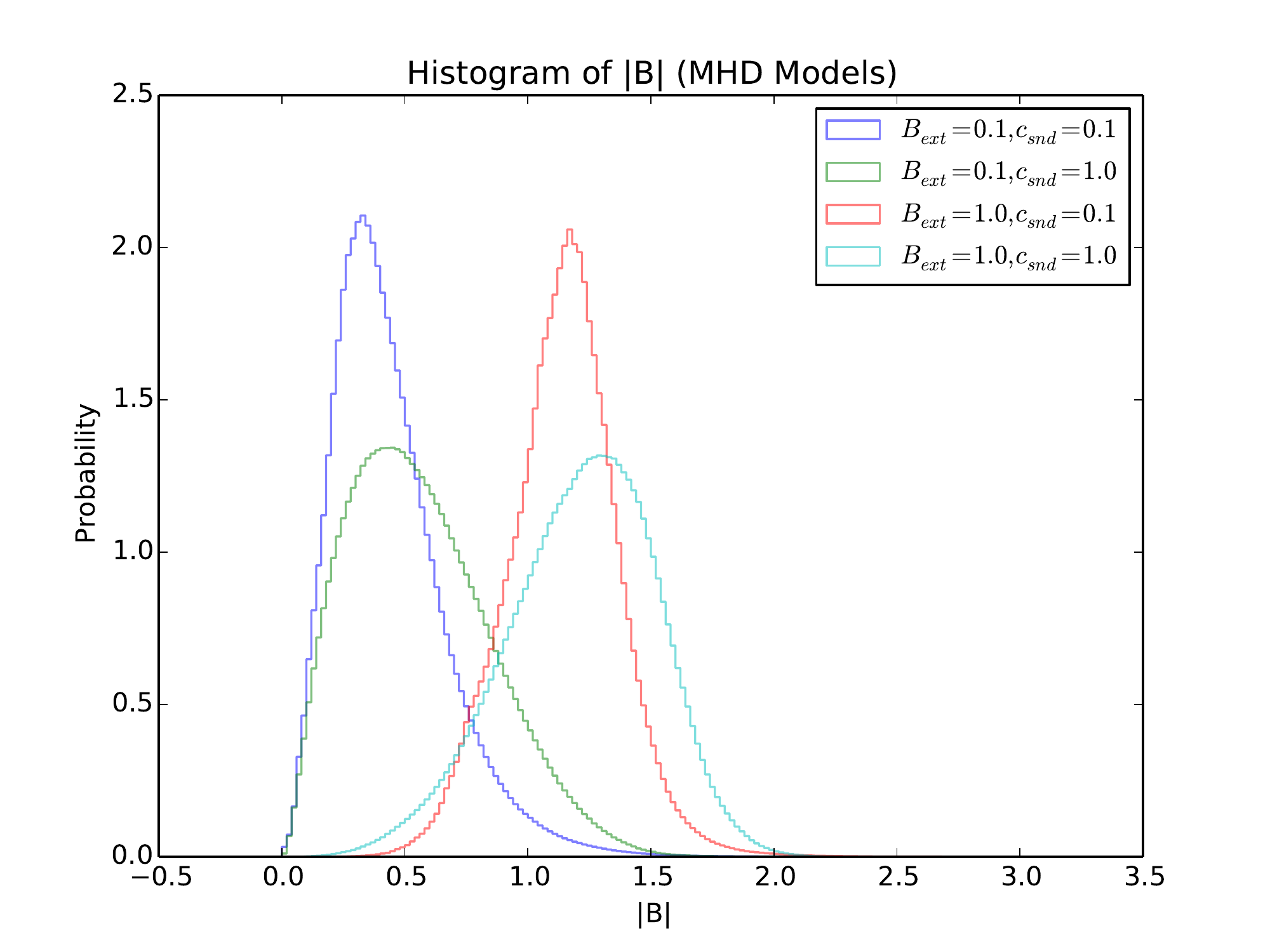}
\includegraphics[width=0.48\textwidth]{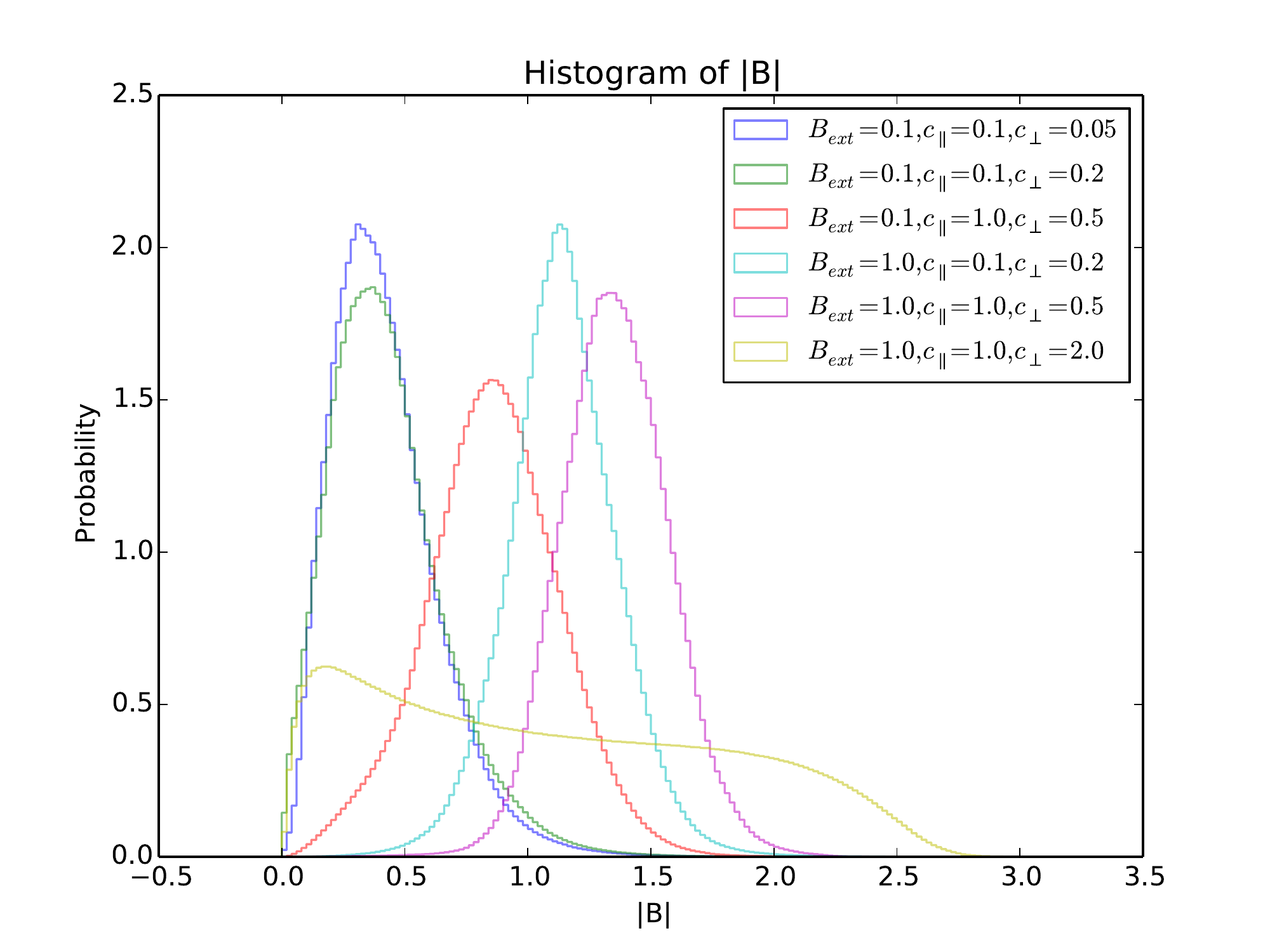}
\caption{Probability density function (PDF) of the magnetic field intensity for
the collisional and collisionless models (left and right panels, respectively).
The models and their parameters are shown in different colors.}
\label{f3h1}
\end{figure*}

\begin{figure*}
\centering
\includegraphics[width=0.48\textwidth]{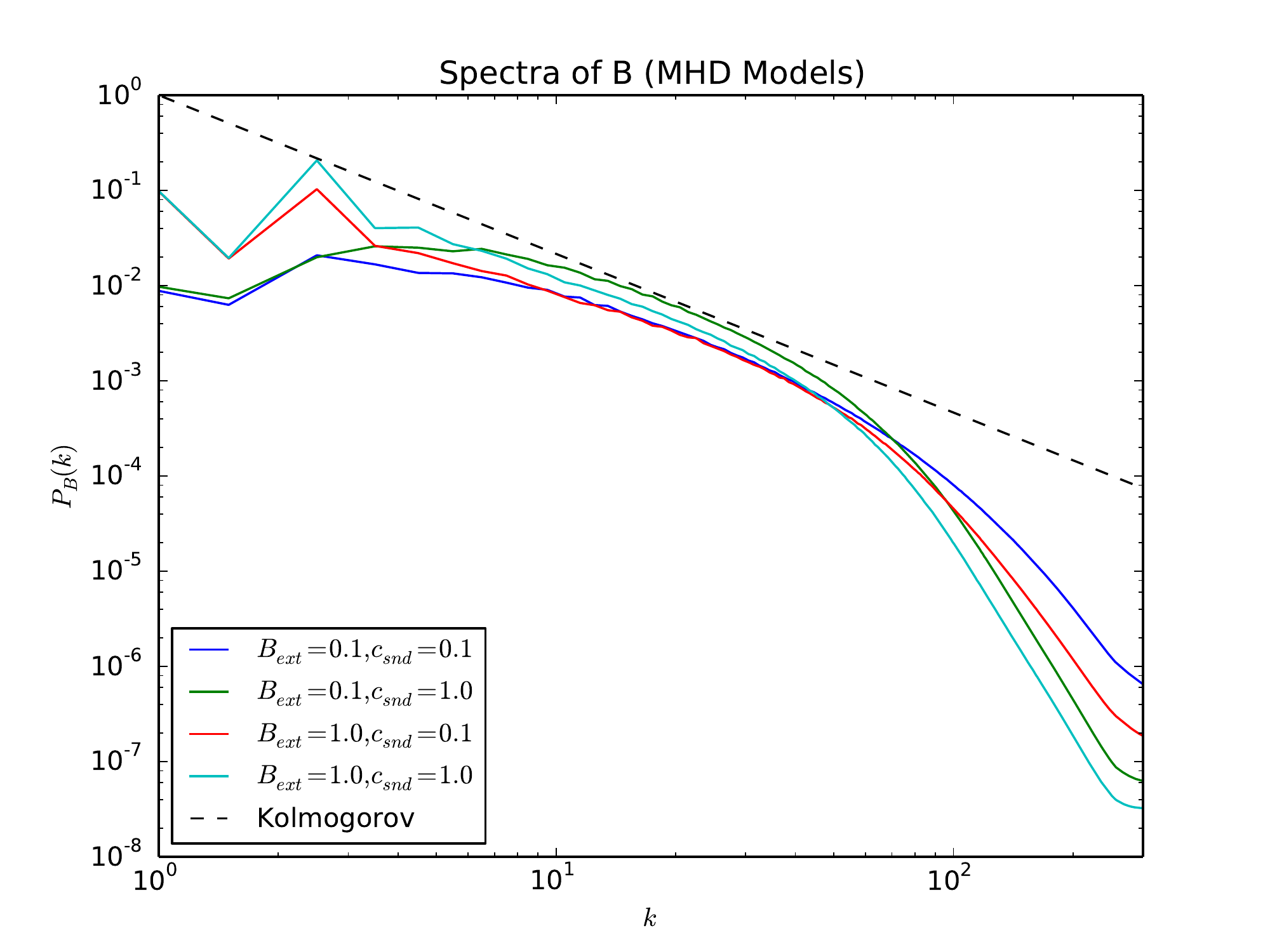}
\includegraphics[width=0.48\textwidth]{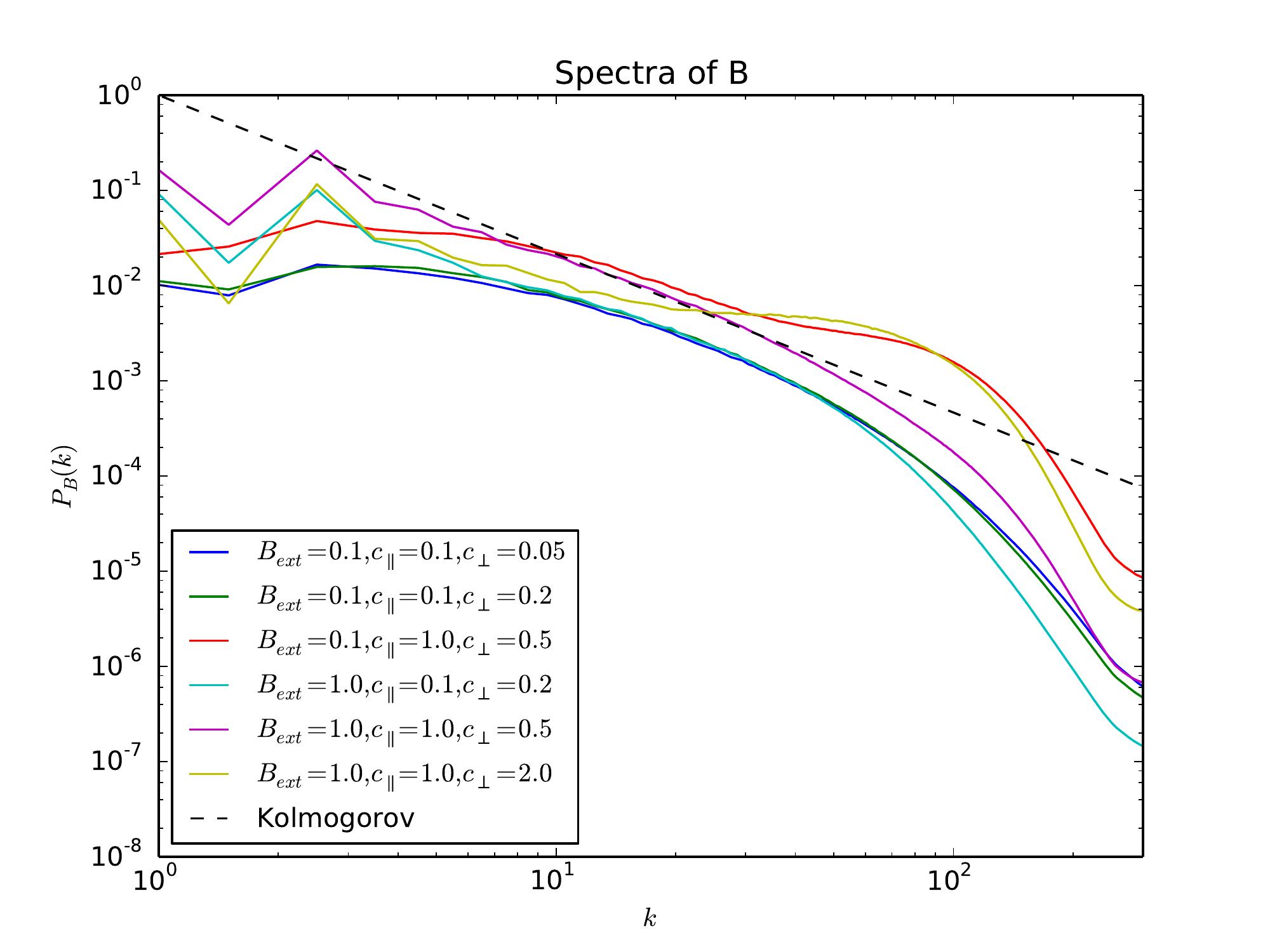}
\caption{Power spectra of the magnetic field for collisional and collisionless
models (left and right panels, respectively). The models and their parameters
are shown in different colors. For comparison, Kolmogorov power spectrum is
shown as dashed black line.}
\label{f2s1}
\end{figure*}

Before analyzing the statistics of the FR maps, it is important to revisit the
main characteristics of the magnetic field that can be extracted from its
probability density function (PDF) and its power spectrum.  The first of these
two quantities retains information on the distribution of the magnetic field
intensity, while the second one on the energy distribution over spatial scales
for each turbulent regime. We show the PDFs of the magnetic field intensity in
Fig.~\ref{f3h1} and the magnetic field spectra in Fig.~\ref{f2s1} summarizing
the main observed features of these two quantities as an extension of the
previous work done by \citet{Kowal_etal:2011}. Additionally, Table \ref{tab:vsk}
lists the statistical moments (variance, skewness, and kurtosis) of the
calculated PDFs in both cases.

We should note that the magnetic field intensity PDFs are not expected to follow
the Maxwellian distribution. This is from a simple fact, that even though the
magnetic field components could in principle follow the Gaussian distribution,
we take into account a uniform component $B_{ext}$ along the X component in our
models. The Maxwellian distribution is expected only if the mean values of the
components are zero. The higher statistical moments are used in this section to
make qualitative comparison between models, and not to determine the fluctuation
randomness or deviation from the Gaussian distribution.

The left panel of Fig. \ref{f3h1} shows the PDF of the magnetic field intensity
for the collisional MHD models. It can be seen that for all the regimes studied
the PDFs have positive skewness for super-Alfv\'enic turbulence (blue and green
lines showing collisional cases corresponding to Models 3-4 and Model 5,
respectively) as well as for the sub-Alfv\'enic turbulence case of Model 6 (red
line) in agreement to the values shown in Table \ref{tab:vsk}. Comparing the
values of $|B|$ corresponding to the distribution maxima, or respectively the
mean values in Table~\ref{tab:vsk}, we see that the distributions are peaked
around the values somewhat larger than the initial value of the field $B_{ext}$
($0.1$ for Models 3--5, and $1.0$ for Models 1--2 and 6). This indicates that
some sort of magnetic dynamo process takes place in those models, either due to
turbulence or kinetic instabilities. For the collisional MHD models it can be
seen from Table \ref{tab:vsk} that the only negative skewness value corresponds
to Model 1-2 (sub-Alfv\'enic and transonic regime). Although this value is
small, the tendency towards left side can still be observed in the corresponding
panel of Fig. \ref{f3h1}. On the contrary, the highest value of skewness is
observed in superAlfv\'enic and supersonic regime with a value reaching over 1.0
and its distribution strongly skewed to the smaller values. Apart from the clear
dependence on the initial value of the mean field in the magnetic field
intensity PDFs, we distinguish from Fig.~\ref{f3h1} that the distributions
depend also on the sound of speed (or sonic regime). Clearly, for supersonic
models the PDFs are more peaked. We notice from both Fig. \ref{f3h1} and Table
\ref{tab:vsk} that the kurtosis of the PDF in the MHD case results either a
leptokurtic for the supersonic regime (red and blue lines) or a platikurtic for
the subsonic regime (cyan and green lines). This could be interpreted as the
presence of infrequent extreme deviations in the values, probably due to the
propagation of pressure waves in the leptokurtic case or plasma waves in the
platikurtic case (see also \citet{Kowal_etal:2011}). It is also interesting to
note that the variance is in all the cases similar being slightly higher for
Model 1-2 (cyan line).

\begin{table}
\centering
\begin{tabular}{crrrr}
\hline\hline
Model & Mean & Std. Dev. & Skewness & Kurtosis\\
\hline\hline
\multicolumn{5}{c}{collisional models} \\
1-2 &  1.215 &  0.302 & -0.311 & -0.056 \\
3-4 &  0.439 &  0.231 &  1.236 &  2.903 \\
5   &  0.562 &  0.287 &  0.569 & -0.039 \\
6   &  1.143 &  0.226 &  0.167 &  1.419 \\
\hline
\multicolumn{5}{c}{collisionless models} \\
1   &  1.091 &  0.706 &  0.306 & -1.052 \\
2   &  1.356 &  0.221 &  0.040 &  0.616 \\
3   &  0.431 &  0.235 &  1.017 &  2.057 \\
4   &  0.421 &  0.217 &  1.128 &  2.682 \\
5   &  0.859 &  0.271 &  0.026 &  0.218 \\
6   &  1.143 &  0.224 &  0.139 &  1.293 \\
\hline
\end{tabular}
\caption{Statistical moments of the magnetic field intensity $|\vec{B}|$ for all models from Table~\ref{tab:models}.} \label{tab:vsk}
\end{table}

The right panel of Fig. \ref{f3h1} shows the PDF for the magnetic field
intensity in the collisionless MHD models. The first information one can extract
from these plots is the volume fraction of the domain where the firehose and
mirror instabilities can develop by observing the volume of the plasma for which
the magnetic field intensity is smaller than the threshold values presented in
Table~\ref{tab:instabilities}. We see that only Models 1, 3, and 5 have a no
negligible fraction of their volume which is unstable.

Similarly to the standard MHD, two cases corresponding to super-Alfv\'enic
turbulence (i.e., Models 3 and 4, green and blue lines, respectively) present
positive skewness with similar kurtosis, being both leptokurtic. These models do
not demonstrate large deviation from their collisional counterparts. Other
models show also positive skewness with much smaller values, however. It is
remarkable that according to the values of kurtosis from Table~\ref{tab:vsk},
only Model 1 (yellow line) has a negative kurtosis, being clearly platikurtic as
shown in Fig. \ref{f3h1}. It is particularly noticeable that the conditions
prevailing in the ICM (Model 5) result in the most Gaussian-like distribution
for the magnetic field intensity with skewness and kurtosis values smaller than
for the other models. It is worth to mention that even though the distribution
is the closest to Gaussian, the peak is shifted to larger $|B|$ as a result of
the firehose instability action leading to larger magnetic field strength
values. For the variance there are no noticeable dissimilarities to the MHD
cases, with a spreader distribution for Model 1.

The spectral distribution of the magnetic fields in the simulations is
quantified by their power spectra shown in Fig. \ref{f2s1}. As mentioned in
Introduction, many observational works assume a random magnetic field with
spectrum following a power law index \citep{Murgia_etal:2004} to link
observational data to synthetic Faraday Rotation maps. Through their
observational method they find, typically, a power law for the magnetic field
spectrum ranging from $-5/3$ to $-11/3$ when considering different galaxy
clusters \citep[see e.g.][]{Murgia_etal:2004, Bonafede_etal:2010b}. Previous
numerical works based on the collisional MHD description of the ICM
\citep{Jones_etal:2011} seem to favor a power law spectrum near $-5/3$, thus
close to Kolmogorov, justifying the use of this value in the construction of
synthetic rotation maps \citep[see e.g.][]{Bonafede_etal:2010}.

The left panel of Fig.~\ref{f2s1} shows the power spectrum for the magnetic
field in the standard MHD cases. All models present similar spectra with a
noticeable inertial range of about one decade in scales and decay at
approximately the same length scales. The amplitude of the power spectrum within
the inertial range is related to the power of the turbulence injected in the
simulations. In terms of the slope of the power spectrum all the MHD models seem
to present a Kolmogorov-like index close to $-5/3$.

In the case of the collisionless MHD models (shown in right panel of Fig.
\ref{f2s1}) the spectral distribution of the magnetic field may be rather
different. Models 2, 3, 4 and 6 (magenta, green, blue, and cyan lines,
respectively) resembles the MHD cases for all spatial scales with a small
departure at larger scales. This reflects the fact that for all these models the
instabilities are not significant (or as in the case of Model 3, are not
dynamically important). However, for Models 1 and 5 (yellow and red,
respectively), which present dynamically important instabilities, there is a
noticeable departure from the standard MHD case at small scales. This departure
manifests as a bump above the dissipation scales. Models 1 and 5, show that both
mirror and firehose instabilities - depending on the turbulent regime - can
produce different spatial distributions for the magnetic field lines which may
impact the statistics of observables, e.g. Faraday Rotation.

\subsection{Effects of temperature anisotropy on synthetic polarization maps}
\label{sec:results:anisotropy_effects}

\begin{figure*}
\centering
\includegraphics[width=\textwidth]{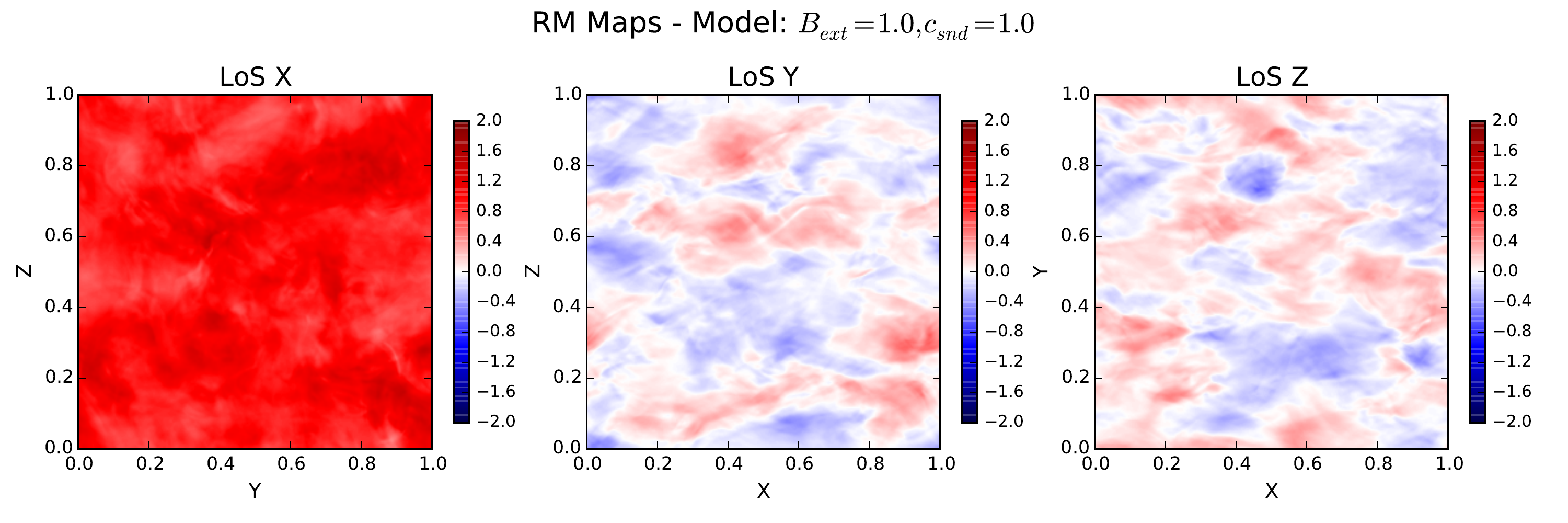}
\includegraphics[width=\textwidth]{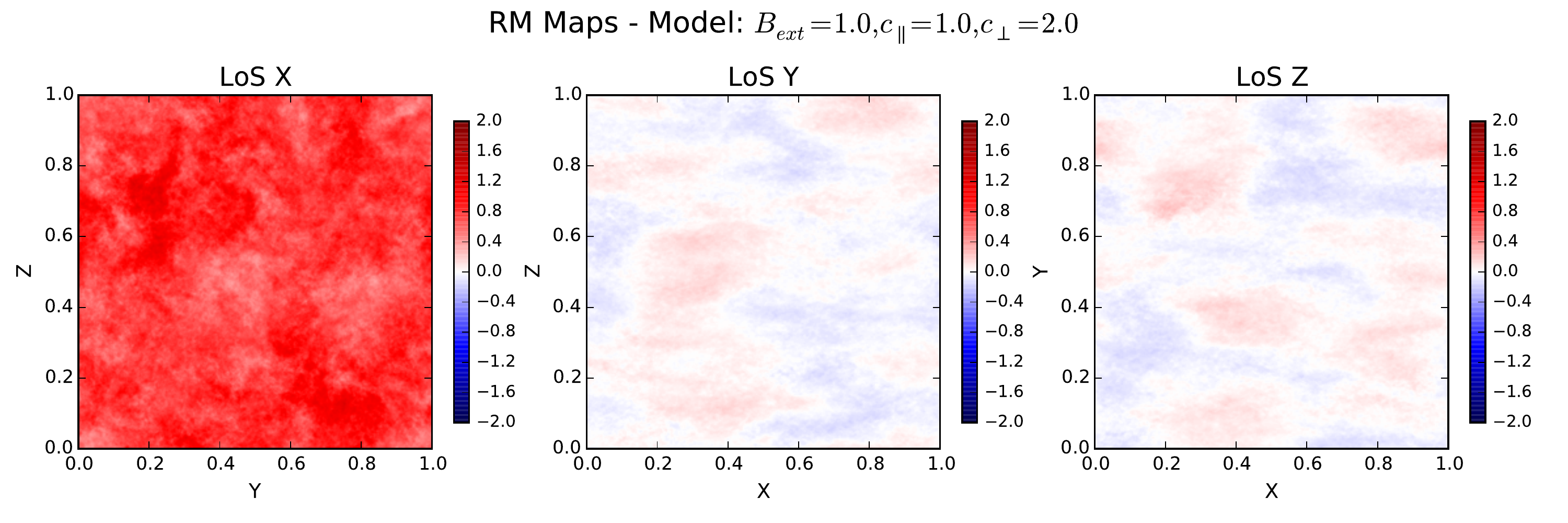}
\includegraphics[width=\textwidth]{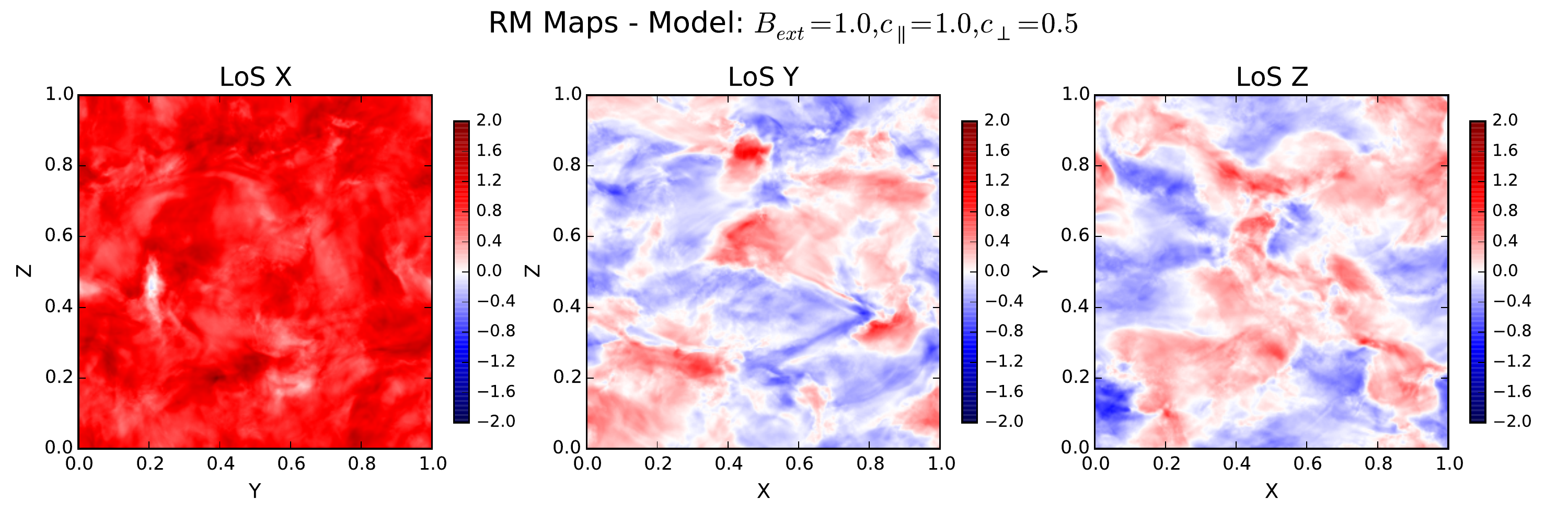}
\caption{Faraday Rotation maps integrated along the $\hat{x}$ axis, which is
parallel to mean field (left column), and perpendicular direction $\hat{y}$ and
$\hat{z}$ (middle and right columns, respectively) for collisional Model 1-2
(upper row) and collisionless Models 1 and 2 (middle and lower rows,
respectively). The maps has the same color scale.}
\label{f4rm1}
\end{figure*}

\begin{figure*}
\centering
\includegraphics[width=\textwidth]{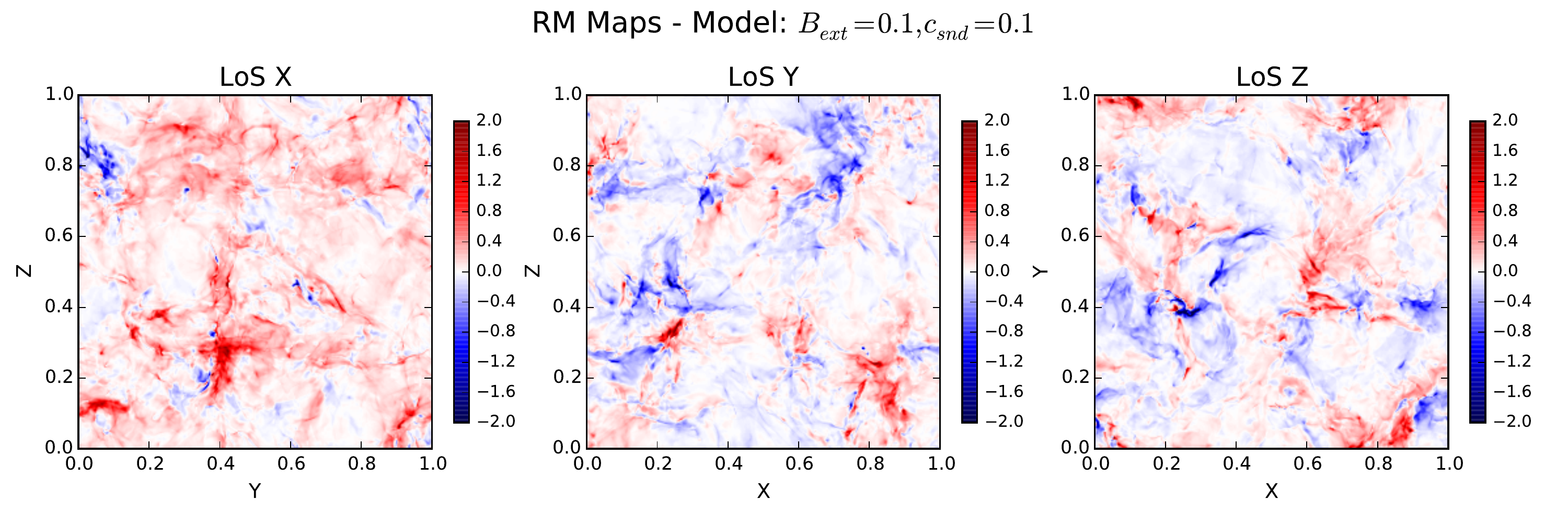}
\includegraphics[width=\textwidth]{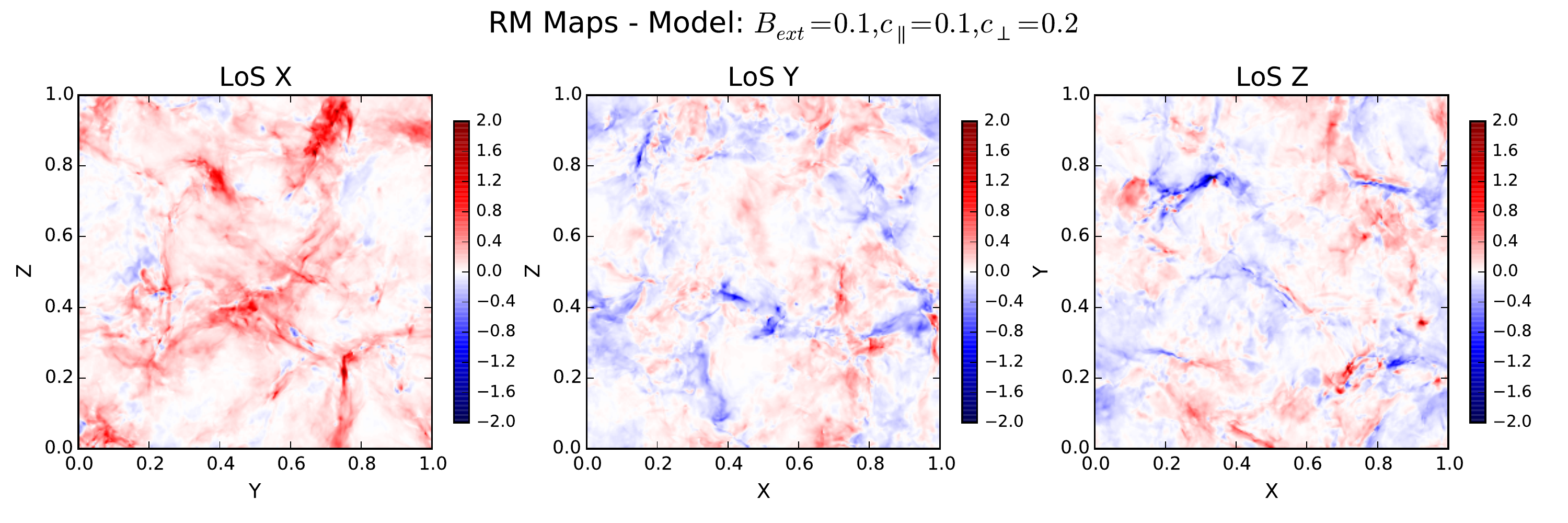}
\includegraphics[width=\textwidth]{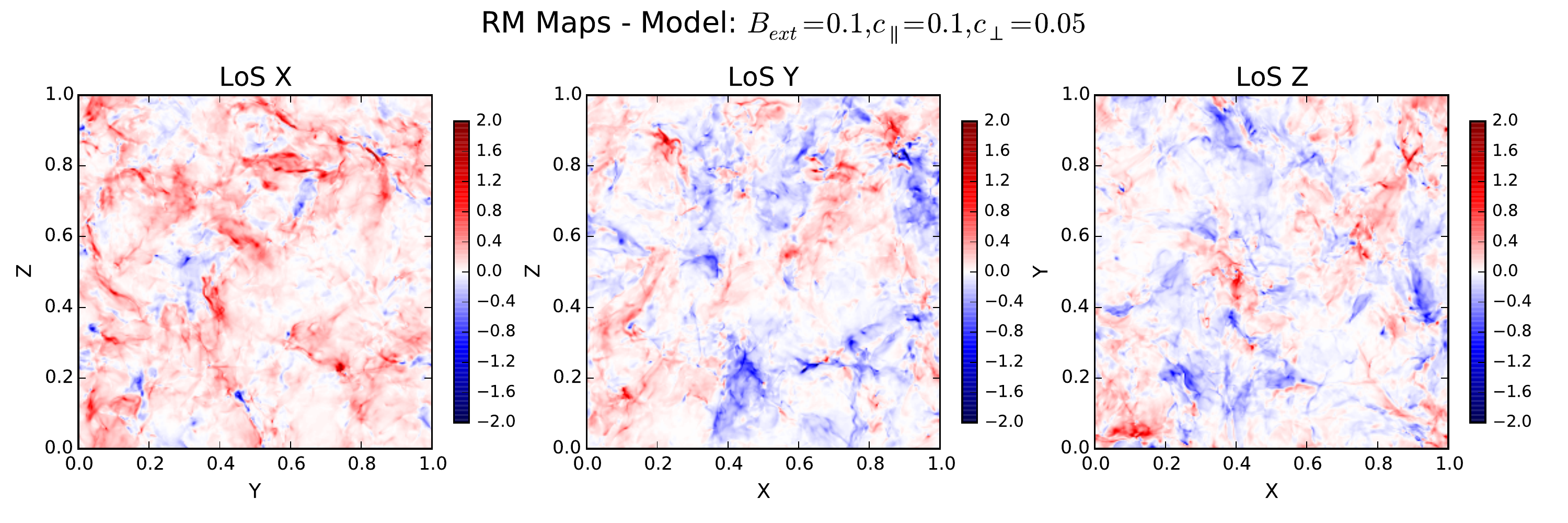}
\caption{Faraday Rotation maps integrated along the $\hat{x}$ axis, which is
parallel to mean field (left column), and perpendicular direction $\hat{y}$ and
$\hat{z}$ (middle and right columns, respectively) for collisional Model 3-4
(upper row) and collisionless Models 3 and 4 (middle and lower rows,
respectively). The maps has the same color scale.}
\label{f4rm2}
\end{figure*}

\begin{figure*}
\centering
\includegraphics[width=\textwidth]{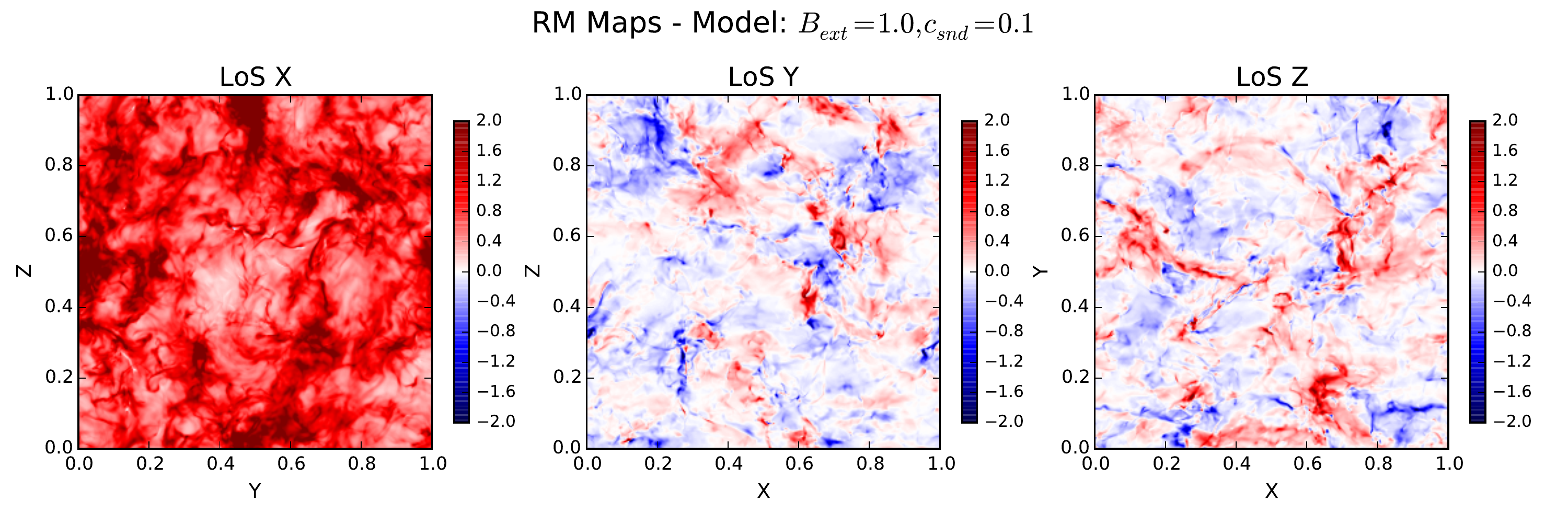}
\includegraphics[width=\textwidth]{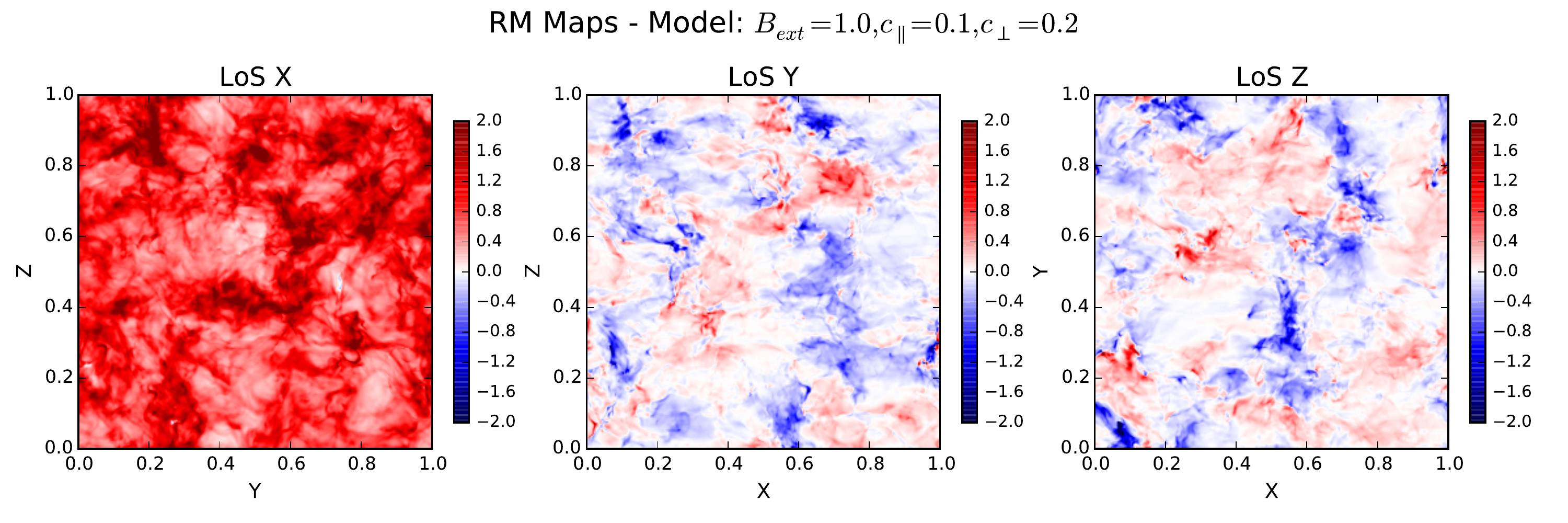}
\caption{Faraday Rotation maps integrated along the $\hat{x}$ axis, which is
parallel to mean field (left column), and perpendicular direction $\hat{y}$ and
$\hat{z}$ (middle and right columns, respectively) for collisional and
collisionless Model 6 (upper and lower rows, respectively). The maps has the
same color scale.}
\label{f4rm4}
\end{figure*}

As we stated previously, one of the main motivations of our work is to determine
the impact of pressure anisotropy, considering a double-isothermal collisionless
MHD approximation, on the Faraday Rotation maps. We produced synthetic Faraday
rotation maps considering two different LoSs, along the $\hat{x}$ (parallel to
the mean magnetic field direction) and along $\hat{y}$ axis (and equivalently
$\hat{z}$ axis, both perpendicular to the mean magnetic field direction) using
the simulated magnetic field at the final snapshot of the simulations. These
maps are shown in Figs. \ref{f4rm1}--\ref{f4rm4}. One of the most important
things to notice from Figs. \ref{f4rm1}--\ref{f4rm4} is that the Faraday
rotation maps reveal anisotropies that are not directly evident in the magnetic
field intensity (RM is much stronger in the parallel direction comparing to the
perpendicular ones, compare left to the middle and right panels of
Figs.~\ref{f4rm1}--\ref{f4rm4} and Figs. \ref{f1b1}--\ref{f1b4}). This is
because, the magnetic field intensity plots show the point contribution at one
particular cut of the computational domain, while the Faraday Rotation map takes
into account the contribution of magnetic field component parallel to LoS
integrated along the computational domain.

Since we are mainly concerned in determining the impact of plasma instabilities
on Faraday Rotation maps mimicking the ICM conditions, we start analyzing Model
5 and then we compare with the other models. We want to stress not only the
effect of such instabilities but also the possibility of having different
initial magnetic field configurations leading to different regimes that can rise
to similarities in Faraday Rotation maps.

\begin{figure*}
\centering
\includegraphics[width=\textwidth]{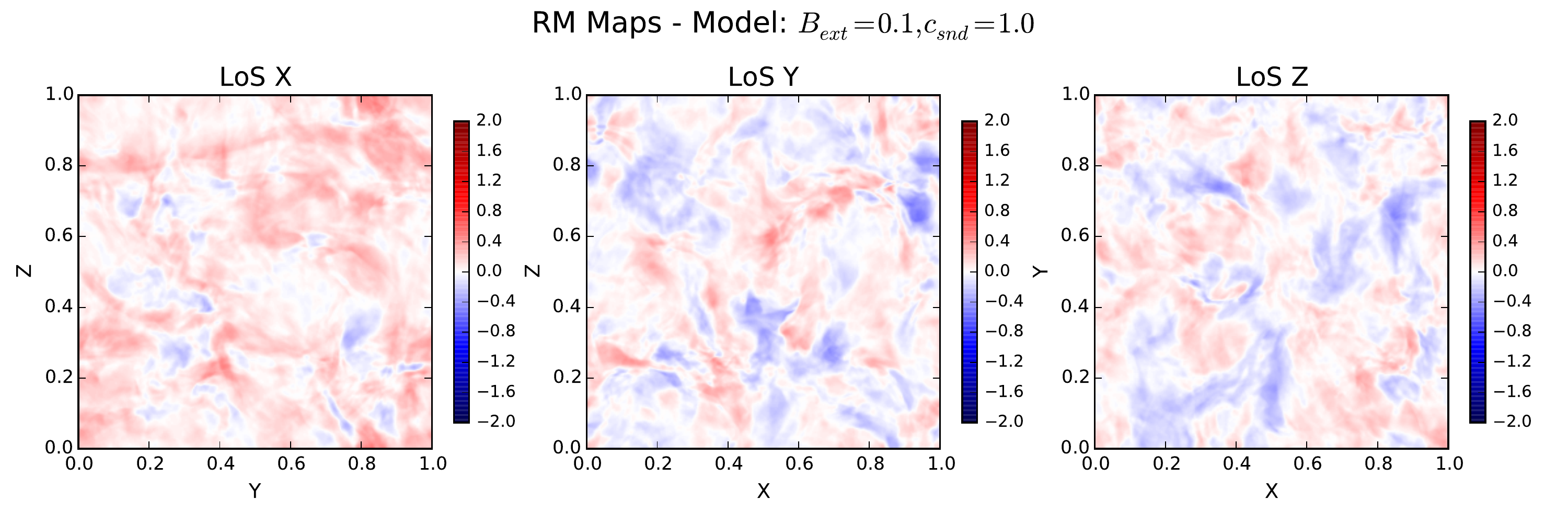}
\includegraphics[width=\textwidth]{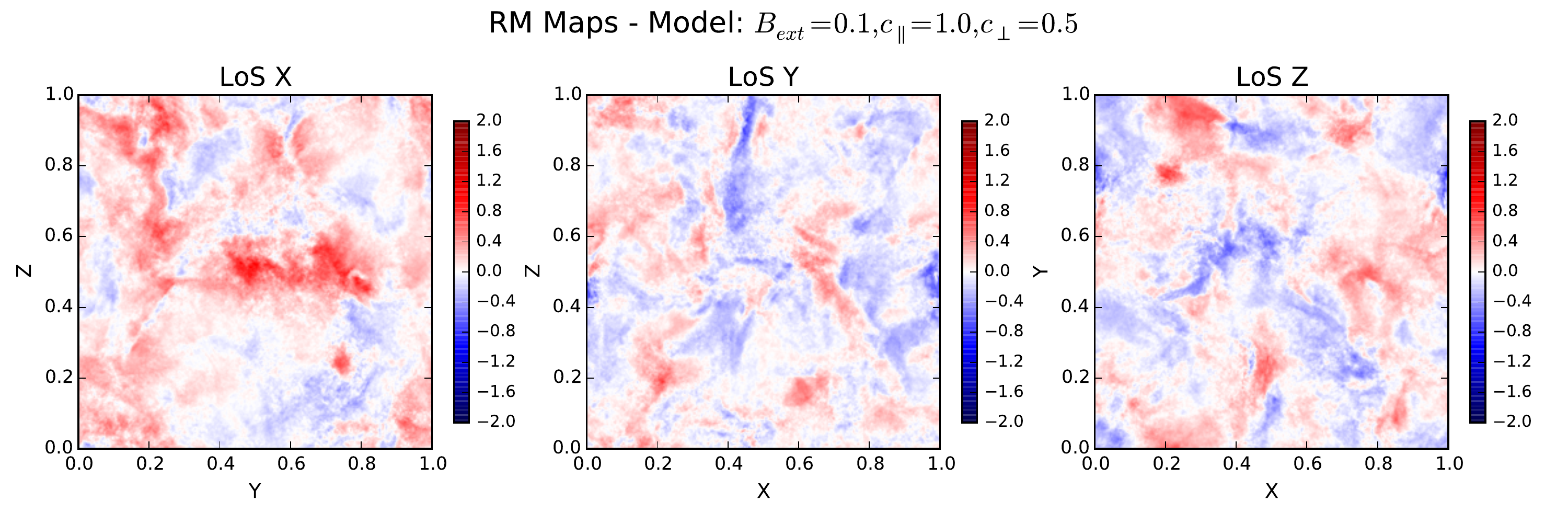}
\caption{Faraday Rotation maps integrated along the $\hat{x}$ axis, which is
parallel to mean field (left column), and perpendicular direction $\hat{y}$ and
$\hat{z}$ (middle and right columns, respectively) for collisional and
collisionless Model~5 (upper and lower rows, respectively). The maps has the
same color scale.}
\label{f4rm3}
\end{figure*}

The Faraday Rotation map of Model 5 (Fig. \ref{f4rm3}), which best resembles the
ICM, presents differences between collisional and collisionless approximations.
In the collisional case it is not likely to reach a robust conclusion for the
initial magnetic field direction due to the resemblance of the maps obtained for
the three LoS. This could be because of the super-Alfv\'enic turbulent regime.
The negative and positive domains are parceled out more or less uniformly in
diffuse filaments with no prevalence of any polarity. The turbulence dominates
the dynamics of the plasma causing a nearly isotropic distribution of the
magnetic field and plasma density which is reflected in the similarity between
the maps. For collisionless MHD, on the other hand, the firehose instability
gives rise to changes in the magnetic field topology dominating over the fluid
motion, but only at small scales. In the LoS parallel to the initial magnetic
field an extended region with higher positive polarity values appears in the
central region of the map. The other two maps present similar distributions with
thicker filaments alternating positive and negative regions. This small scale
effect gives the appearance of more granulated maps, in comparison to the
collisional MHD ones which appear smoother due to the bigger polarization
structures of the maps. In terms of the variance of the maps, these results are
in agreement with the higher variance values obtained for the collisionless MHD
case.

In the case of strong mean field configuration of Model 1 (for collisional and
collisionless MHD, see top and middle panels of Fig.~\ref{f4rm1}) the RM
distribution for parallel direction is mostly positive and much stronger than
along the perpendicular direction to $B_{ext}$. The map features are clearly
different to the ICM case (Model 5). In addition, due to the initial
configuration the maps with a LoS parallel to the initial magnetic field show an
almost random distribution of positive values. This effect is supported by the
smaller values of statistical moments, namely skewness and kurtosis (see Table
\ref{tab:vskrm}). In the case of Model 2, the map corresponding to parallel LoS
differs to the previously discussed cases due to the presence of near zero value
regions which reflects the appearance of patterns in this map. Still, this
behavior is not compatible with a ICM-like feature. The variance values are
higher in this case than those corresponding to the collisional case and to
Model~2.

In the super-Alfv\'enic turbulence regime, i.e. in Models 3 and 4 (see Fig.
\ref{f4rm2}), the first thing to notice is the fact that in collisional as well
as collisionless MHD are not remarkably different. The reason for this effect is
the super-Alfv\'enic turbulence resulting in isotropization of the magnetic
fields.  When the instabilities are impelled due to the pressure anisotropy
(middle and right rows of Fig. \ref{f4rm2}), the distribution of
positive-negative structures on the Faraday maps is diffuse and with the
presence of an average polarity dominated region and tiny filaments of opposite
polarity. For Model 3 (see bottom row of Fig. \ref{f4rm2}), the super-Alfv\'enic
turbulence struggles against the mirror modes which propagate nearly
perpendicular to the magnetic field. In the case of the stable Model 4, no clear
differences between the directions are present. This is supported by similar
values for variance being minimal in the perpendicular direction for Model~3
(see Table \ref{tab:vskrm} and middle panel of Fig. \ref{f4rm2}). Taking into
consideration these results, which depart from those obtained for ICM, it would
be unexpected to find Faraday maps with such characteristics in environments
like the ICM, even though these three models represent cases of super-Alfv\'enic
turbulent regime.

Finally, Fig. \ref{f4rm4} shows the Faraday Rotation maps obtained for
turbulence in supersonic regime. In consistency with the analysis of Model 6 in
the previous section, we see that there is no significant differences between
the collisional and collisionless cases, only that in the $\hat{x}$ LoS the
values of the Faraday maps are larger and near randomly distribuited. This is
because most of the domain in the collisionless model is stable as supersonic
turbulence suppresses the instabilities. The filamentary distribution in the
maps corresponding to the LoS in the perpendicular directions are clearly
well-defined. The spread of the filaments is in agreement with the variance
values of Table \ref{tab:vskrm} being the highest among all models.

\subsection{Probability distribution function and power spectra of the Faraday Rotation maps}
\label{sec:results:rm_pdfs_spectra}

\begin{figure*}
\centering
\includegraphics[width=0.48\textwidth]{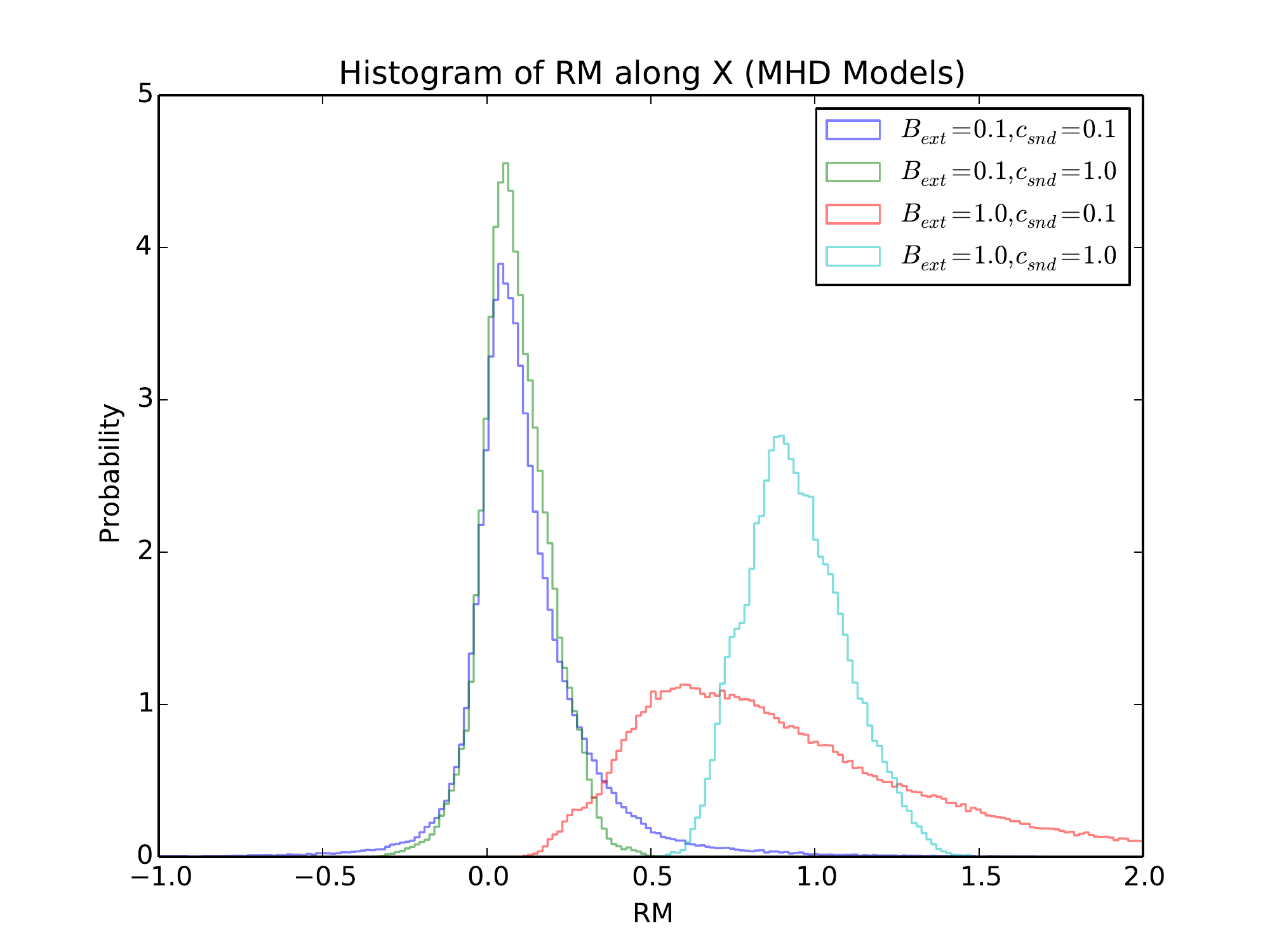}
\includegraphics[width=0.48\textwidth]{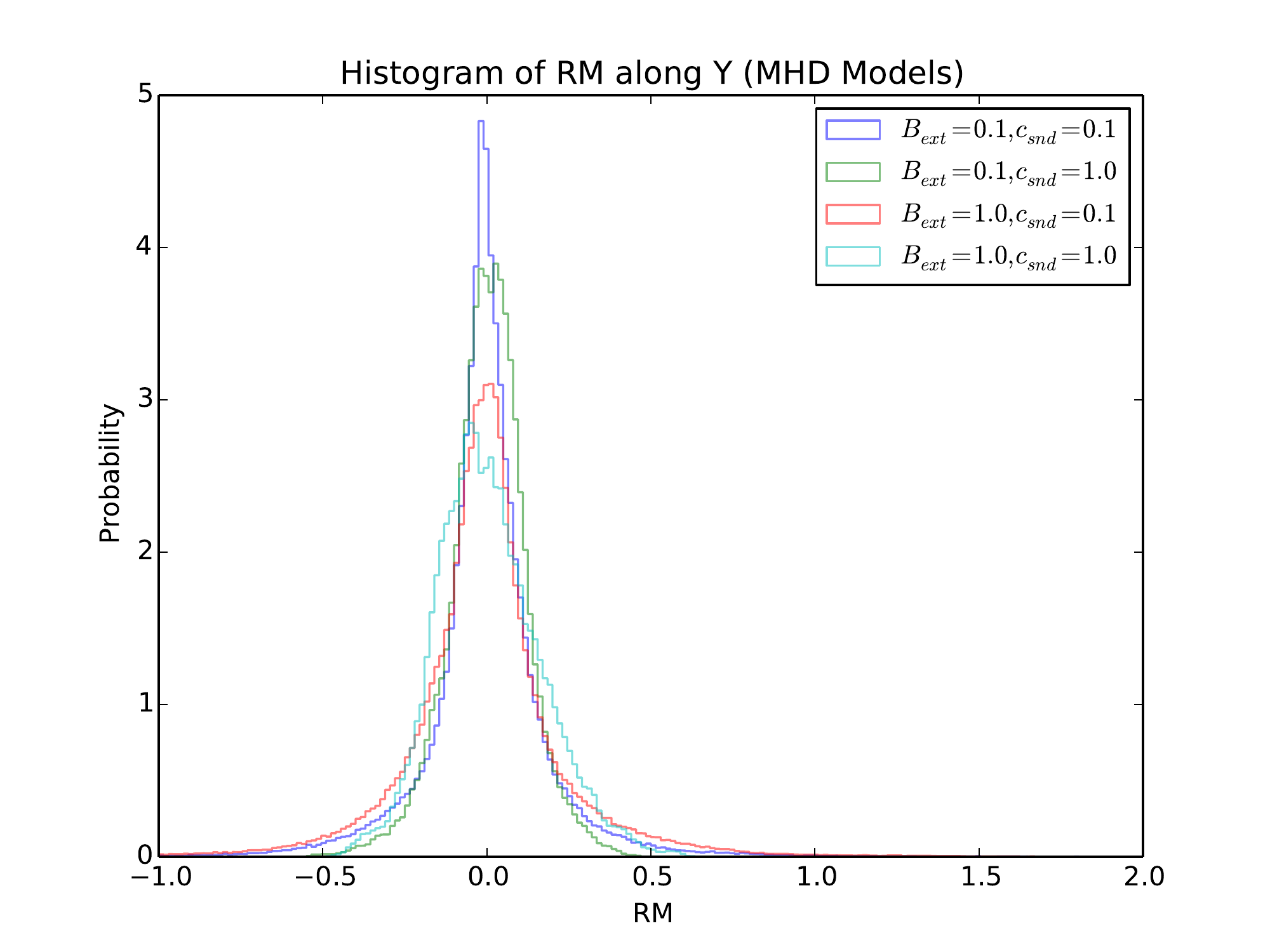}
\includegraphics[width=0.48\textwidth]{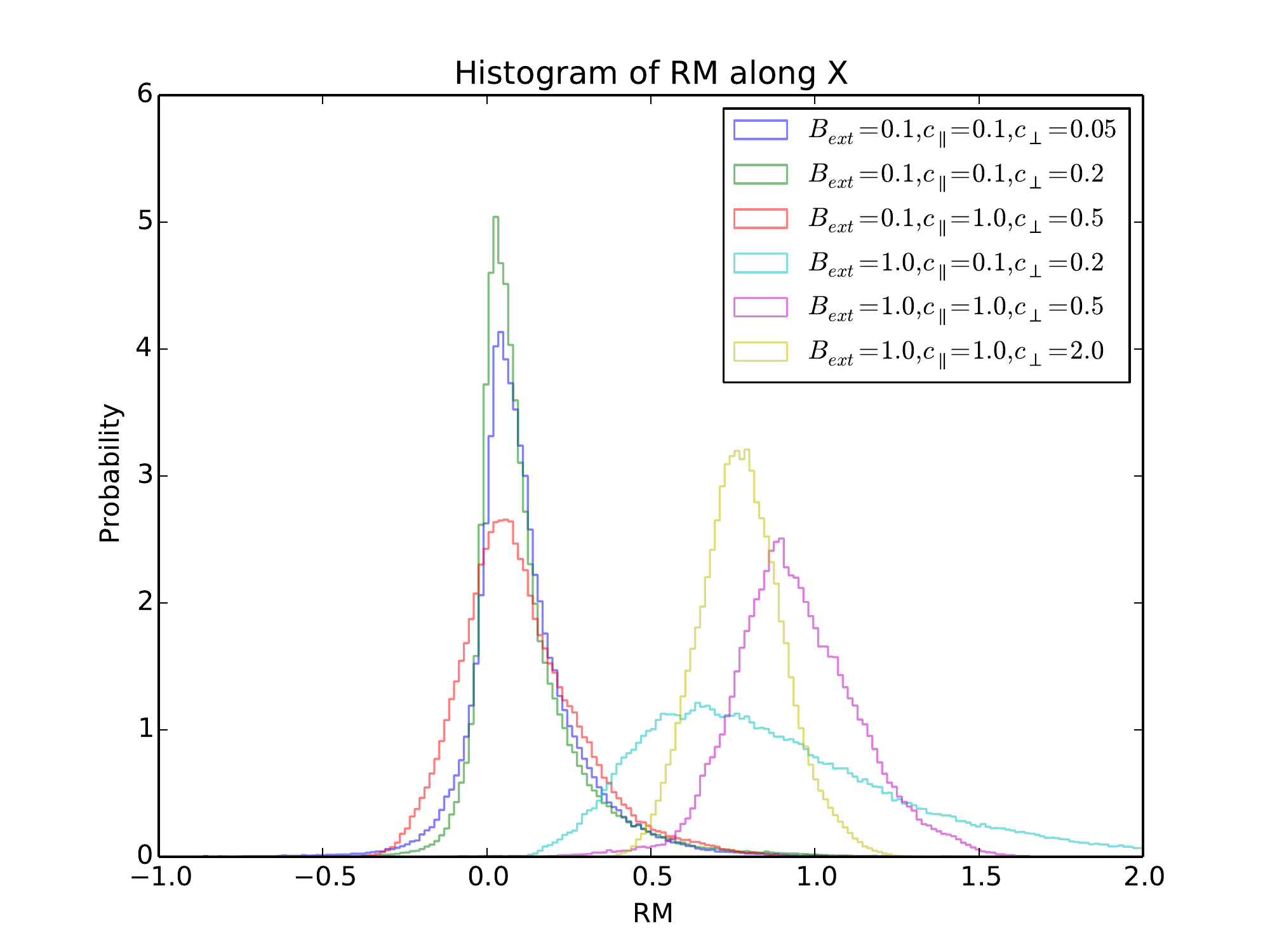}
\includegraphics[width=0.48\textwidth]{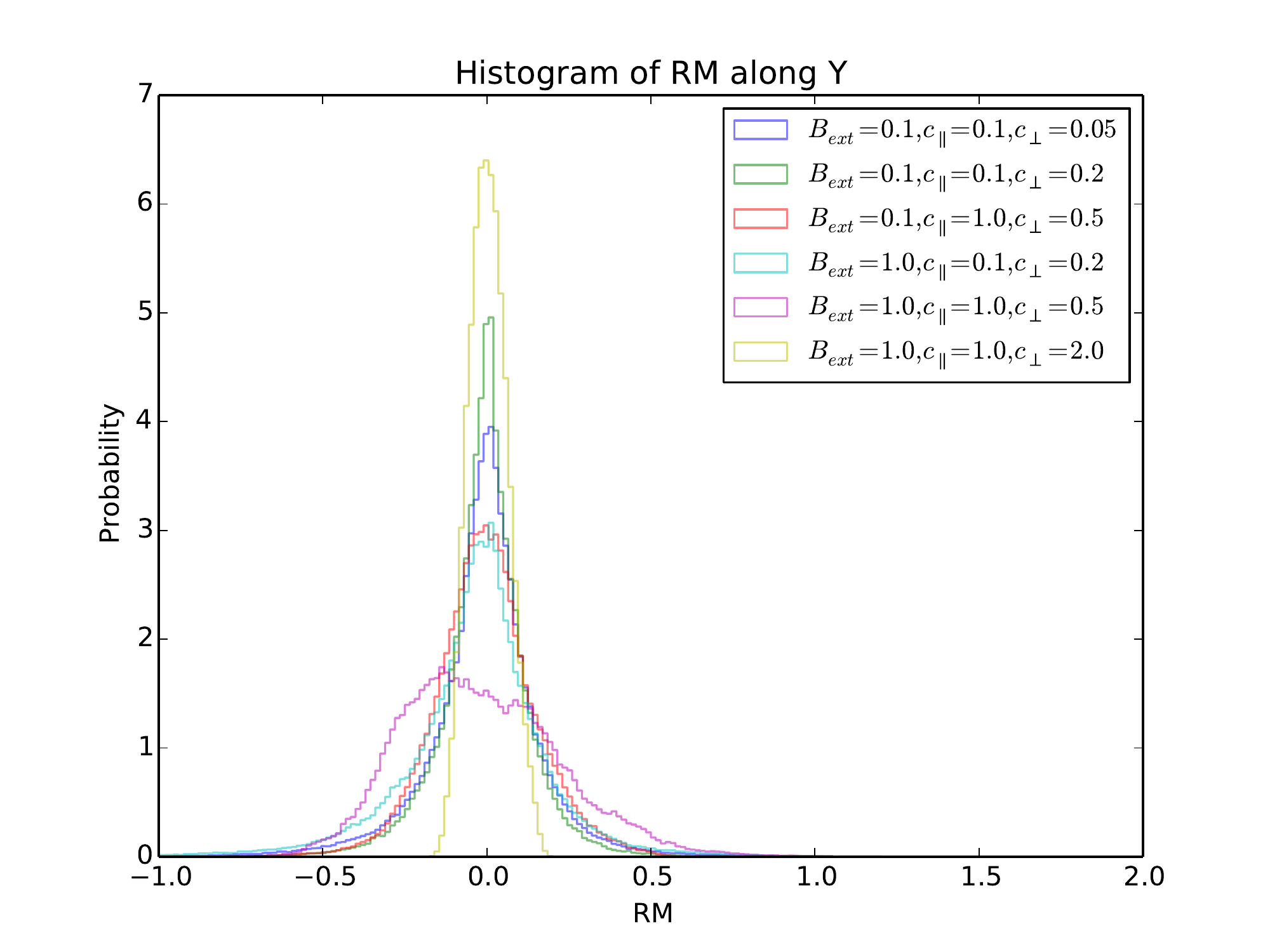}
\caption{Probability distribution function of the RM maps corresponding to the
collisional and collisionless models (upper and lower rows, respectively). Left
and right columns show the RM integration along the parallel and perpendicular
direction to $B_{ext}$, respectively.}
\label{f3h2}
\end{figure*}

As mentioned in Introduction, observational works are compelled to assume a
Gaussian distribution for the magnetic field component along the LoS used to
compute synthetic Faraday Rotation. This is the simplest assumption one can
adopt. Here we investigate synthetic RM maps obtained from self-consistent
magnetic field distributions from three-dimensional numerical simulations of
turbulence which allow for direct derivation of their statistical properties,
such as the probability distribution function and the power spectrum. The PDF of
RM for each model can be seen in Fig. \ref{f3h2} (upper panels for collisional
and lower panels for collisionless models) and the statistical moments of the RM
distributions are given in Table \ref{tab:vskrm}.

\begin{table}
\centering
\begin{tabular}{ccrrrrrr}
\hline\hline
Model & LoS & Mean & Std. Dev. & Skewness & Kurtosis\\
\hline\hline
\multicolumn{6}{c}{collisional models} \\
1-2& $\parallel$ &  0.944 &  0.150 &  0.323 & -0.248 \\
{} & $\perp$     &  0.006 &  0.160 &  0.386 &  0.342 \\
3-4& $\parallel$ &  0.116 &  0.201 &  0.976 & 14.000 \\
{} & $\perp$     & -0.001 &  0.198 &  0.451 & 17.063 \\
5  & $\parallel$ &  0.091 &  0.104 &  0.202 &  0.571 \\
{} & $\perp$     &  0.005 &  0.119 & -0.231 &  1.238 \\
6  & $\parallel$ &  0.971 &  0.559 &  2.050 &  7.802 \\
{} & $\perp$     & -0.006 &  0.253 &  0.180 &  9.522 \\
\hline
\multicolumn{6}{c}{collisionless models} \\
1  & $\parallel$ &  0.777 &  0.130 &  0.207 &  0.075 \\
{} & $\perp$     &  0.005 &  0.058 &  0.163 & -0.356 \\
2  & $\parallel$ &  0.942 &  0.193 &  0.083 &  1.101 \\
{} & $\perp$     & -0.019 &  0.242 &  0.466 &  0.336 \\
3  & $\parallel$ &  0.117 &  0.169 &  2.359 & 10.580 \\
{} & $\perp$     & -0.006 &  0.139 & -0.355 &  5.743 \\
4  & $\parallel$ &  0.112 &  0.175 &  1.218 & 10.100 \\
{} & $\perp$     & -0.009 &  0.185 & -0.677 & 15.303 \\
5  & $\parallel$ &  0.105 &  0.185 &  0.838 &  1.195 \\
{} & $\perp$     & -0.004 &  0.156 &  0.005 &  1.075 \\
6  & $\parallel$ &  0.897 &  0.433 &  1.287 &  2.868 \\
{} & $\perp$     & -0.036 &  0.232 & -0.682 &  7.288 \\
\hline
\end{tabular}
\caption{Statistical moments for the RM maps integrated along the parallel ($\parallel$) and perpendicular ($\perp$) directions to the mean field for all models from Table~\ref{tab:models}.}
\label{tab:vskrm}
\end{table}

The synthetic RM maps are obtained from our cubes by integrating the product of
density and magnetic field component parallel to the integration direction,
i.e.:
\begin{equation}
RM = \frac{1}{L} \int_0^L {\rm d}l \, \rho (l) B_n (l),
\end{equation}
where $L$ is the size of the simulated cube, $l$ is the position along the
integrated direction (our LoS), and $B_n$ is the magnetic field component
parallel to the integration direction (or normal to the plane of map). With this
definition, the mean value of RM over the plane of sky is given by
\begin{equation}
\overline{RM} = \rho_0 \bar{B}_{n} + \frac{1}{L} \overline{ \int_0^L {\rm d}l \, \delta\rho(l) \delta B_n(l)},
\label{eq:rm_avg}
\end{equation}
where $\bar{B}_n$ is the mean of the magnetic field component $B_n(l)$, $\rho_0$
is the average density (equal to unity in all our simulations), $\delta B_n(l) =
B_n(l) - \bar{B}_n$ is the fluctuating part of magnetic field components, and
the bar denotes the average over the plane of sky. The last term in
Eq.~\ref{eq:rm_avg} is zero if both the density and magnetic field fluctuations
have normal distributions. In such case, $\overline{RM} = \rho_0 \bar{B}_n$, and
e.g. if the integration is done along the $\hat{x}$ direction, $\overline{RM} =
B_{ext}$, or if the integration is done along the $\hat{y}$ or $\hat{z}$
directions, $\overline{RM} = 0$. In the next paragraphs we will see, that the
mean values of $RM$ deviate from these values, indicating that the distributions
of $\delta \rho$ and $\delta B_n$ are not described by Gaussian distributions.
The spread (or width) of the RM distribution $\sigma_{RM}$ can be approximately
given by
\begin{equation}
\sigma^2_{RM} \sim \delta B^2_{f} + \bar{B}_n \delta \rho^2_{f},
\end{equation}
where $\delta B^2_f$ and $\delta \rho^2_f$ are the power of the magnetic and
density fields at the scales of the energy injection.

Analyzing the mean $RM$ for collisional models in Table~\ref{tab:vskrm}, we see
that for the LoS perpendicular to the mean magnetic field, $\overline{RM}
\approx 0$ (the difference is always smaller than $0.01$), while for the
parallel LoS, it is smaller than $B_{ext}$ for Models 1-2, 5, and 6 (6\%, 9\%,
and 3\% smaller, respectively) and larger for Model~1-2 (16\% larger). In the
parallel RM maps of the collisional models (top left panel of Fig.~\ref{f3h2}),
subsonic models (green and cyan lines) present normal-like distribution of RM.
Two subsonic models demonstrate deviation from the normal distribution in the
high-value tail (Model~6 shows extremely elongated tail justifying the large
value of skewness in Table~\ref{tab:vskrm}). This is most probably due to the
log-normal distribution of the density in the highly supersonic regime
\cite[see]{Kowal_etal:2007}. Both super-Alfv\'enic models do not differ much,
only the less peaked form of the curve characterizes Model~5, due to the
mentioned log-normal distribution of density. For the perpendicular LoS, we see
in the right top panel of Figure \ref{f3h2} that the width of the distribution
of the RM measurements are all similar (see the standard deviation column in
Table~\ref{tab:vskrm}). For the parallel LoS, models with low $B_{ext}$ show
similar spread, even though Model 3-4 (blue line) is supersonic, for which the
contribution of the density fluctuations is of minor importance as it is
multiplied by small $B_{ext}$. However, when comparing models with larger
$B_{ext}$ (Models 1-2 and 6, cyan and red lines in Figure~\ref{f3h2},
respectively), we clearly see that the spread of the RM distribution of Model~6
is much larger due to the supersonic turbulence \cite[also, see the comparison
between both density power spectra in][]{Kowal_etal:2011}.

Now, lets focus on how the collisionless models deviate from their collisional
counterparts. First, we notice from Table~\ref{tab:vskrm}, that Model~1 is much
lower mean value of $RM$ than its collisional counterpart. This model has RM
distribution relatively symmetric, however, strongly shifted to the lower values
(see yellow line in bottom left panel). This must be the effect of strong mirror
instability operating here. The mean $RM$ for Models 2--5 do not differ much
comparing with their collisional counterparts, including Model 5 which best
describes the ICM. Model~6 demonstrate lowered mean value of RM, although its
distribution is similar to its collisional counterpart (see cyan line in the
left lower panel and red line in the left upper panel, for collisionless and
collisional cases, respectively). For perpendicular LoS (the right bottom panel
in Figure~\ref{f3h2}), the values of $\overline{RM}$ are around zero (see also
Table~\ref{tab:vskrm}), similarly to the collisional case. Only Models 1 and 6
show mean values slightly higher than others in this direction. The width of the
RM distributions, however, for Model~1 (yellow line) is smaller, while for
Model~2 (magenta line) is larger. Other models do not show significant
differences with respect to their collisional counterparts, even Model~5
representing the ICM. With regard to the width $\sigma_{RM}$ of the
distribution, we observe its reduction for Models~1 and 3 (yellow and green
lines, respectively) while it increases for Models~2 and 5 (magenta and red
lines). It is easy to understand these differences in terms of the changes in
the effective Alfv\'en speed (which increases for Model~1 and 3 and decreases
for Models~2 and 5).

For perpendicular LoS, it can be seen in Figure \ref{f3h2} that all curves,
except collisionless Model~2, lie almost in the same region as the collisional
models. All plots being similarly symmetric and leptokurtic. The RM distribution
of collisionless Model~2 is strongly deformed.

Now, lets evaluate the higher order statistical moments of the distributions.
For parallel LoS, all the models in both collisional and collisionless cases
lead to positive skewness values with a minimum value corresponding to
collisionless Model~2, which is much smaller than in its collisional counterpart
(see Table~\ref{tab:vskrm}). Model~5, representing the ICM conditions, has
relatively large value of skewness, much higher than the corresponding
collisional model. Similarly, collisionless Models 3 and 4 have larger
skewness when comparing to their collisional counterpart. In Model~6, the
skewness of distribution decreases after including pressure anisotropy.

Looking at the kurtosis column in Table~\ref{tab:vskrm}, we see that for the
perpendicular maps, most of the distributions are strongly peaked (kurtosis
$>1.0$). The Model~1 is slightly platykurtic, while Model~2 is slightly
leptokurtic. The ICM Model~5 has kurtosis $~1.1$ in the direction perpendicular
to mean field, which is not much different for the corresponding collisional
case. The biggest difference in distribution between collisional and
collisionless cases in the perpendicular map is observed for Model~2, which is
much more spread comparing to its collisional counterpart.

\begin{figure*}
\centering
\includegraphics[width=0.48\textwidth]{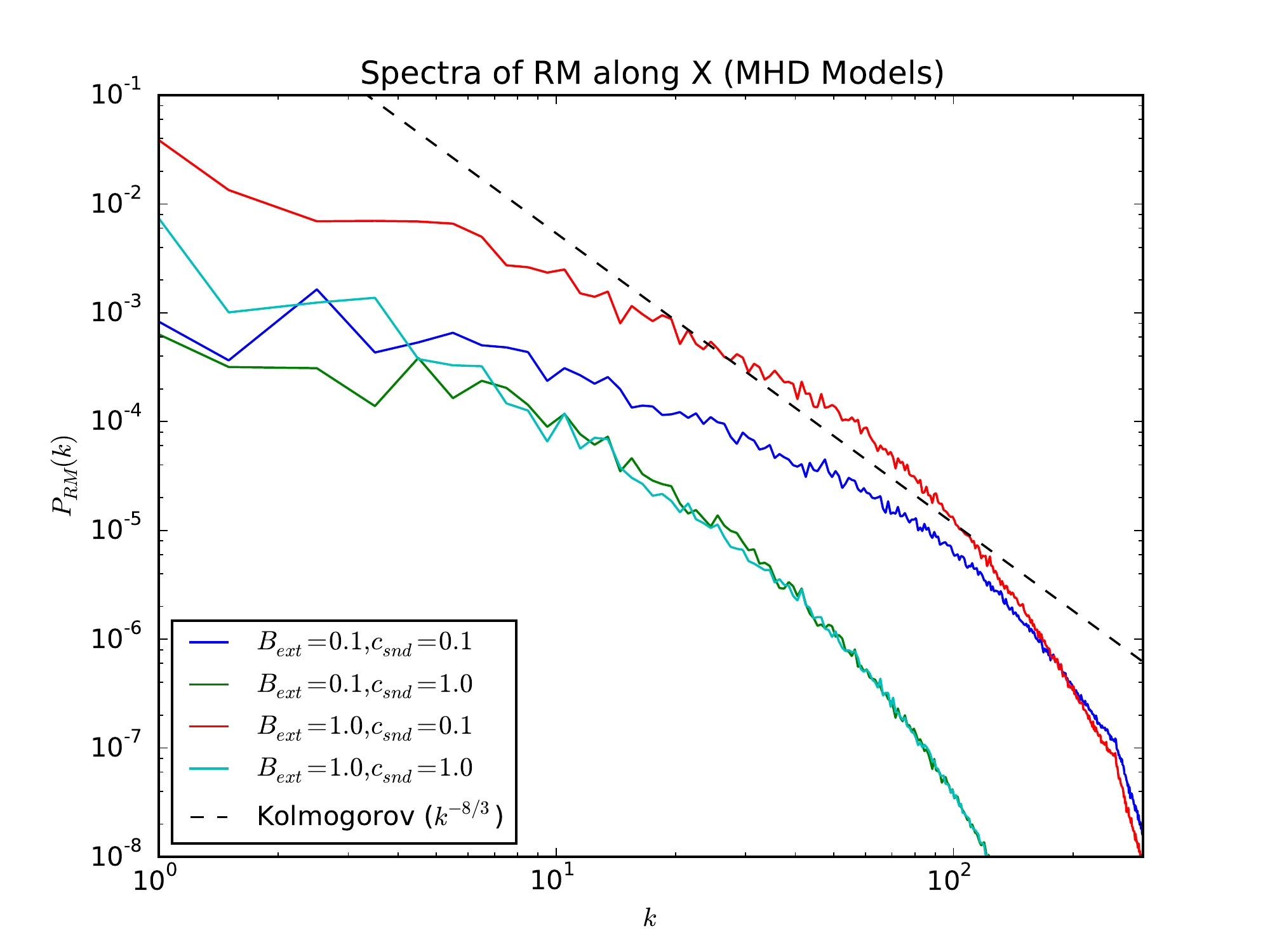}
\includegraphics[width=0.48\textwidth]{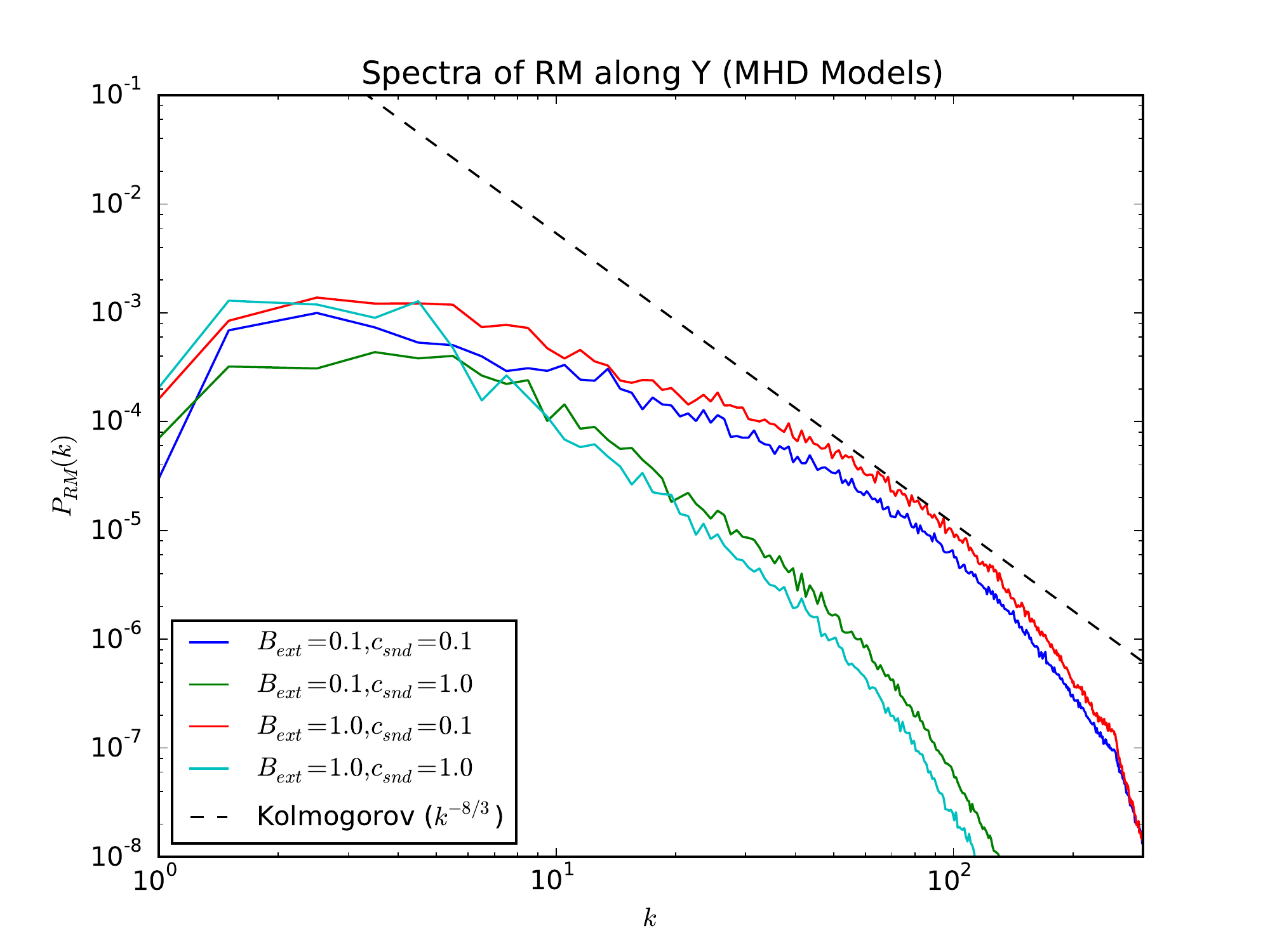}
\includegraphics[width=0.48\textwidth]{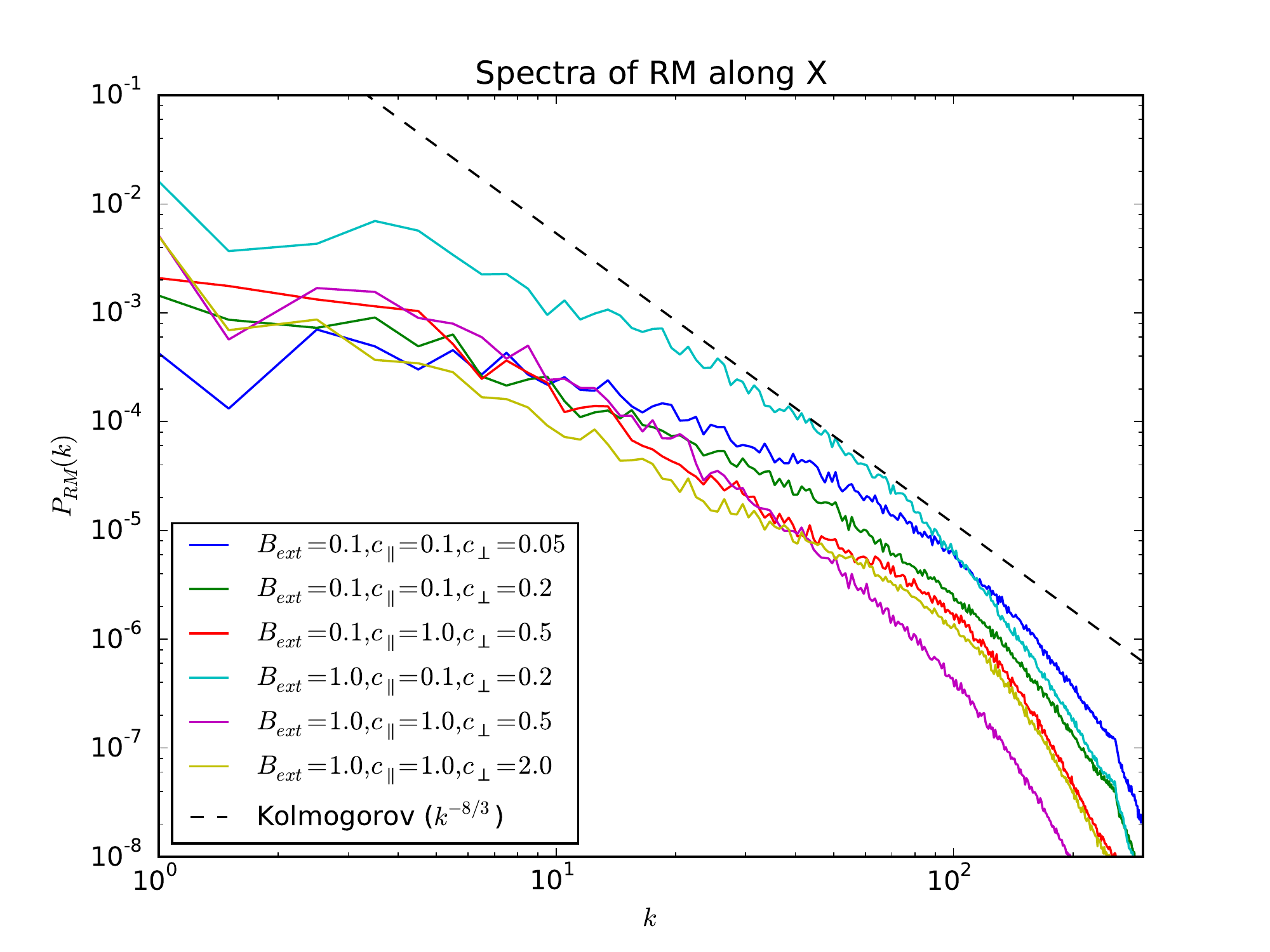}
\includegraphics[width=0.48\textwidth]{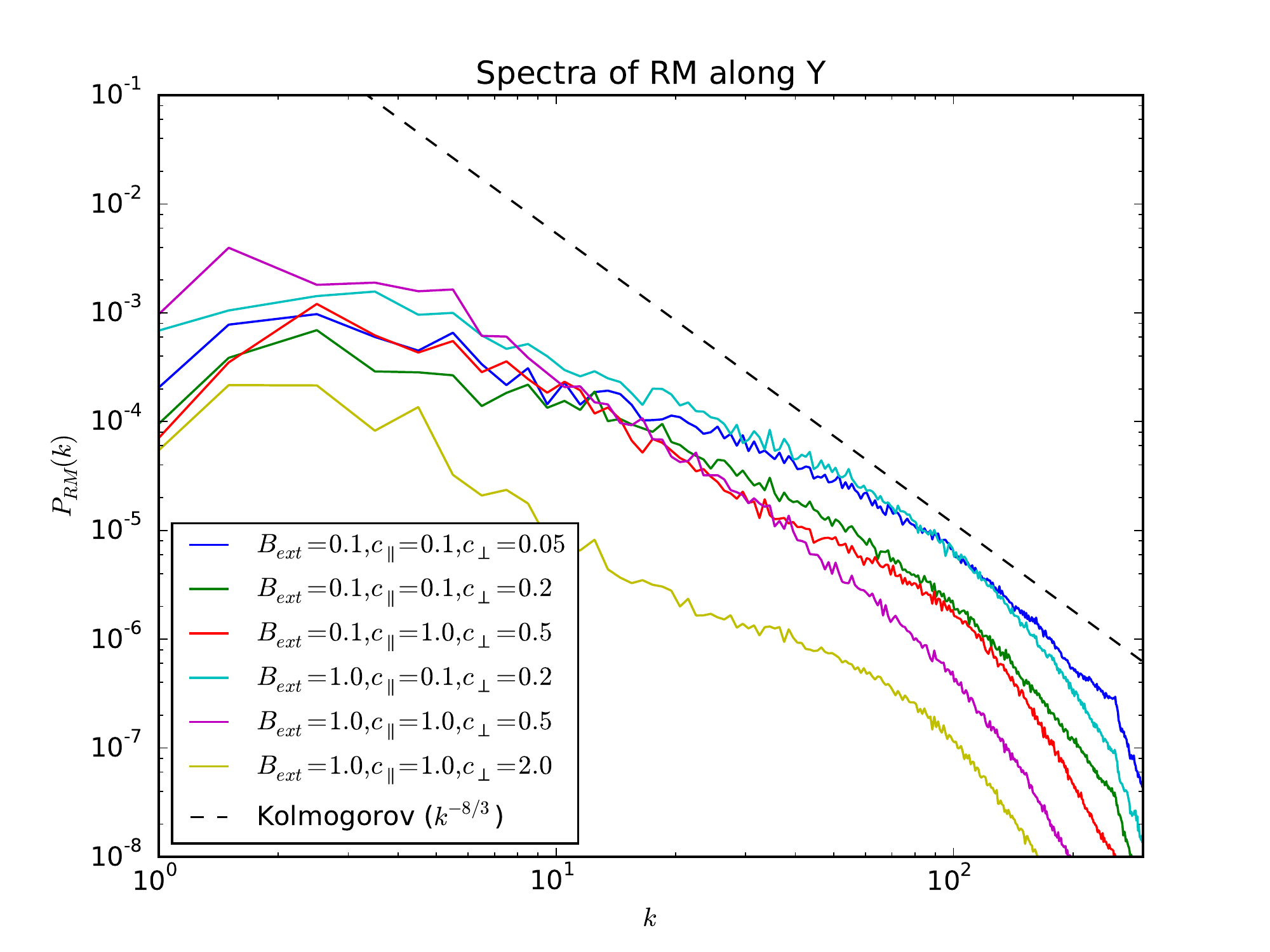}
\caption{Power spectra of RM for collisional and collisionless models (upper and
lower panels, respectively). Spectra for RM integrated along the parallel and
perpendicular direction to $B_{ext}$ are shown in the left and right columns,
respectively. The 2D Kolmogorov power spectrum ($k^{-8/3}$) is shown in dashed
black line for comparison.}
\label{f2s2}
\end{figure*}

The power spectra of RM obtained for each model is shown in Figure \ref{f2s2}.
For collisional models (upper panels) the remarkable is the distinction between
both the transonic (Models 1-2 and 5) and supersonic regimes (Models 3-4 and 6)
(cyan-green versus blue-red lines, respectively) for both directions.

In the limit of anisotropic turbulence and neglecting any magnetic-density phase
correlation, the RM power spectrum can be related to the density and magnetic
power spectra in the following way:
\begin{eqnarray}
{\cal P}_{RM} (k) \propto
\left( \rho_0^{2} \frac{P_B(k)}{k} + \bar{B}_{n}^{2} \frac{P_\rho(k)}{k}
\right).
\end{eqnarray}

In the case of the perpendicular LoS ($\bar{B}_{n} = 0$ ), the estimative above
states that only the magnetic power spectrum influences ${\cal P}_{RM}(k)$. In
the transonic cases (green and cyan lines of Figure \ref{f2s2}, upper panel),
the slopes seem to be compatible with $k^{-8/3}$ (as it is expected if
$P_B(k)\sim k^{-5/3}$). In the supersonic cases, we have also the contribution
of magnetic fluctuations coming from the magnetosonic modes. As the magnetic
spectrum of the fast modes is flatter \citep[$\sim k^{-3/2}$ according
to][]{ChoLazarian:2002}, we could expect the magnetic spectrum to be flatter.
Indeed, we see such feature in the magnetic power spectrum of Fig. ~\ref{f2s1},
but it is not enough to explain the difference seen in the RM spectra (which
satisfies $\sim k^{-1}$). Therefore, such difference must comes from the
contribution of the combined density and magnetic fluctuations from the
compressible modes in the smaller scales (larger $k$ values). Comparing with the
parallel LoS case, the subsonic models (cyan and green lines) have slopes
similar to the perpendicular LoS. In these cases, the density fluctuations are
small and do not affect the ${\cal P}_{RM}(k)$. Also for the supersonic and
super-Alfv\'enic Model 3-4 (blue line), parallel LoS does not differ from the
perpendicular LoS case. This is due to the fact that the contribution from the
density fluctuations are weighted by the $\bar{B}_{n}$, which is small
($B_{ext}=0.1$) in this case. Only the sub-Alfv\'enic, supersonic Model 6 (red
line) differs in the parallel LoS. Here, the contribution from the density
fluctuations seems to increase the power in the large scales, although the slope
seems steeper than in the perpendicular LoS.

Now, lets examine the differences in the ${\cal P}_{RM}$ for the collisionless
models show in the lower row of Figure \ref{f2s2}. The ICM representative
Model~5 (red line) looks similar to its collisional counterpart in large scales,
however, in the small scale range the fluctuations of $RM$ are stronger. The
similar situation is seen in the perpendicular LoS (the bottom right panel). It
is due to the enhanced magnetic fluctuations originating from the firehose
instability. The transonic Model 2 (violet) and Models 4 and 6 are also similar
to its collisional counterpart in both LoSs. For the perpendicular LoS, Model 1
(yellow line) presents less power in most of the the scales, compared to the
collisional case. Also, at the large scales, the power law seems to be deeper.
The decrease (increase) of magnetic power in the large (small) scales (see
Fig~\ref{f2s1}) cannot explain such differences, so it must be a combined effect
of the density-magnetic field fluctuations (which have negative correlation for
mirror modes). For the parallel LoS, however, we can observe an increase of the
power only at the small scales. It can be due to the increase of the density
power at the same scales \cite[see][]{Kowal_etal:2011}.

\subsection{Autocorrelation functions of Faraday Rotation maps}
\label{sec:results:autocorrelation}

\begin{figure*}
\centering\includegraphics[width=0.48\textwidth]{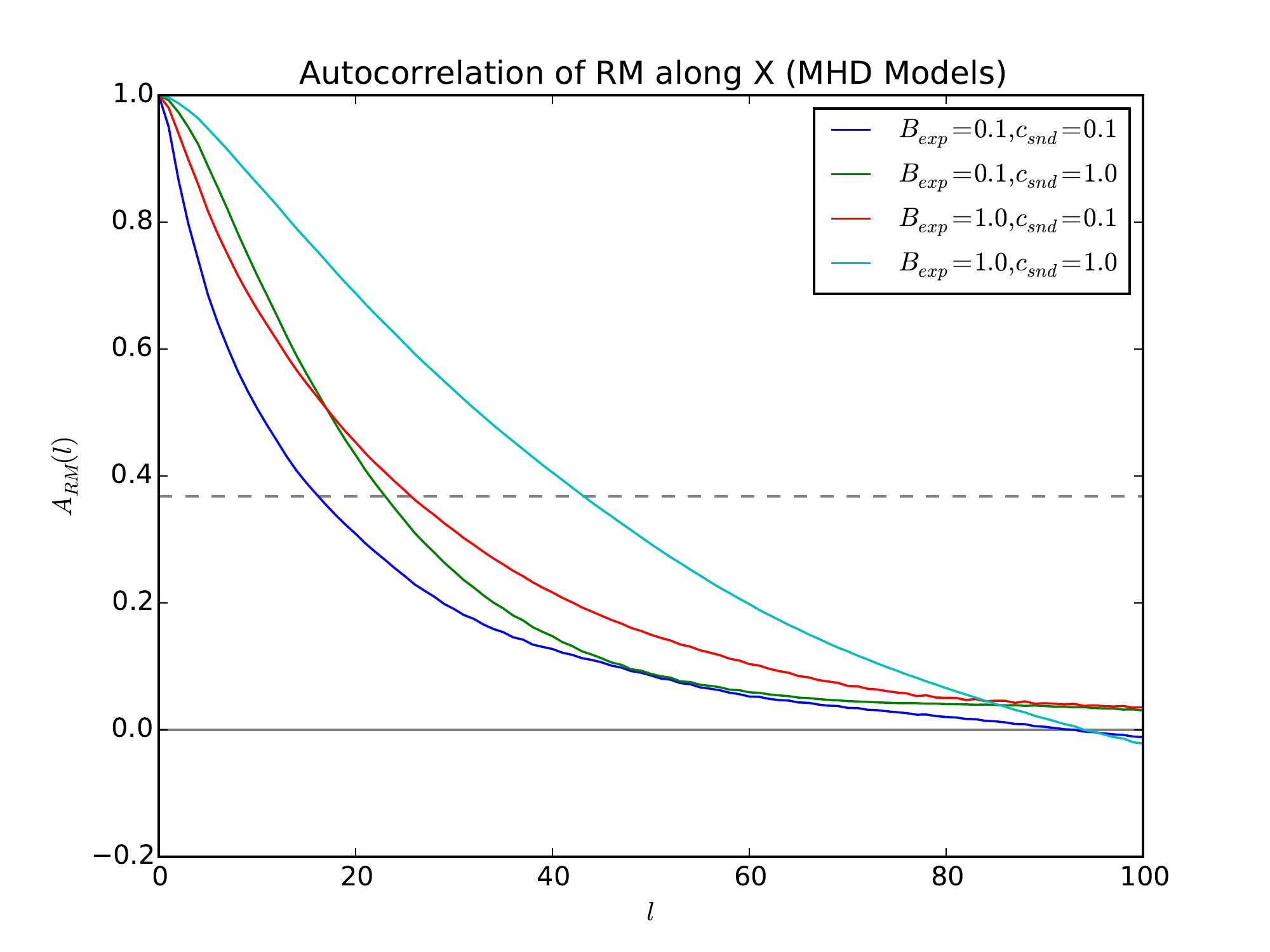}
\centering\includegraphics[width=0.48\textwidth]{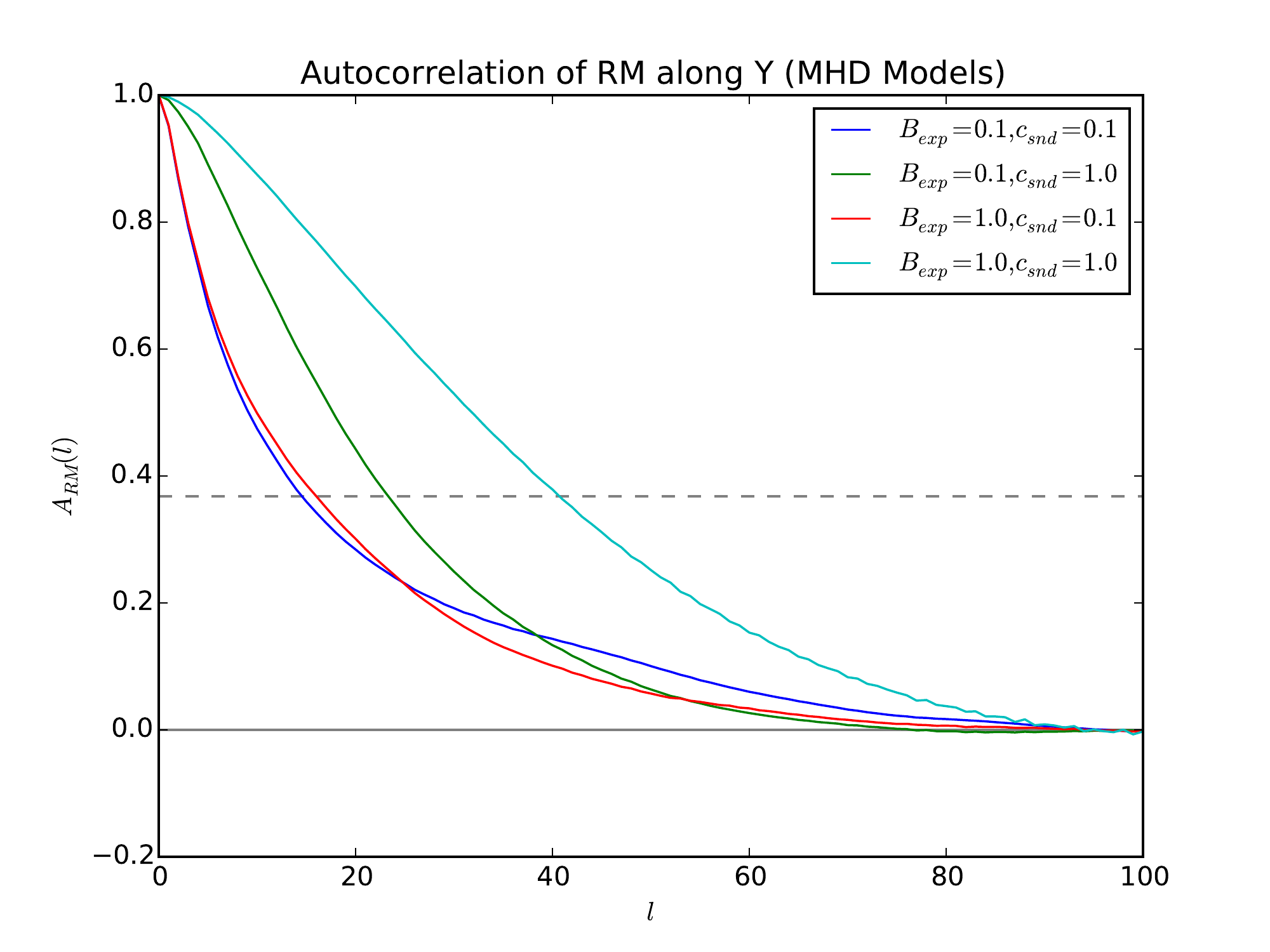}
\centering\includegraphics[width=0.48\textwidth]{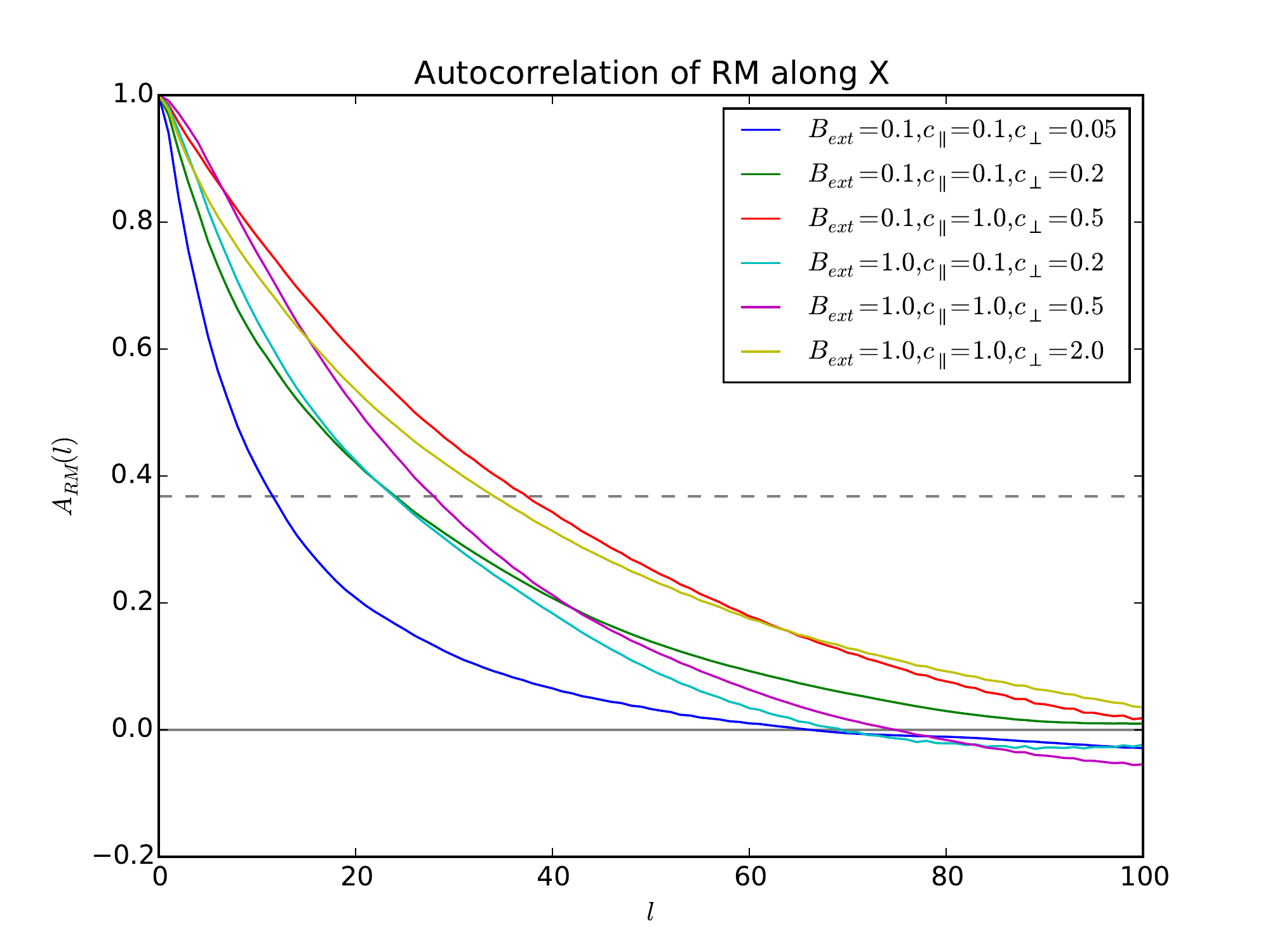}
\centering\includegraphics[width=0.48\textwidth]{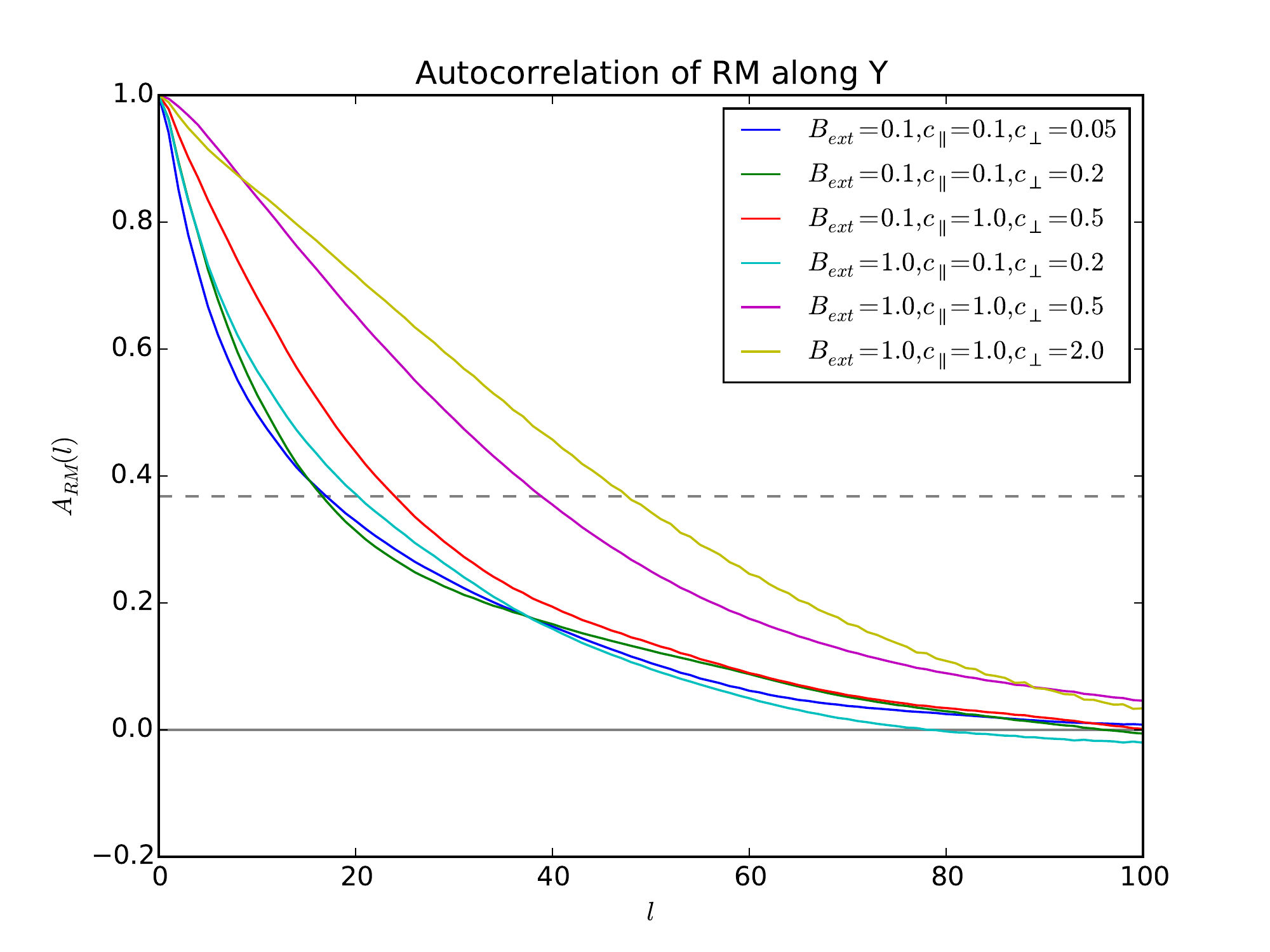}
\caption{Autocorrelation function calculated from the Faraday Rotation maps for
LoS along $\hat{x}$ and $\hat{y}$ (left and right panels, respectively)
corresponding to the collisional and collisionless MHD models (upper and lower panels, respectively)
as a function of the length scale $l$.}\label{f6af1}
\end{figure*}

We compute the isotropic autocorrelation function of the simulated FR maps as
follows:
\begin{eqnarray}
A(l) = \langle RM(\vec{r}) RM(\vec{r} + \vec{l}) \rangle,
\label{acorr}
\end{eqnarray}
where $\langle \cdots \rangle$ stands for the average taken over all positions
$\vec{r} = (x, y)$ in the FR maps and directions of the shift $l = |\vec{l}|$.
It can be related to the power spectrum of $RM$, ${\cal P}_{RM}(k)$, in the
plane by
\begin{eqnarray}
A(l) \propto \int_0^{\infty} {\rm d}k \, {\cal P}_{RM}(k) \cos (kl),
\label{srf1a}
\end{eqnarray}
where $k$ is the wave vector \citep[see a more detailed analysis
in][]{EnsslinVogt:2003}. The use of this function enables us to quantify the
statistical properties of the magnetic field structure. In particular, it is
possible to extract an average correlation length, which gives an idea of the
extension of patches in the structure of the field. This quantity can be
observed in Fig. \ref{f6af1}, which shows the collisional cases in the upper
panels and the collisionless ones in the lower panels.

In subAlfv\'enic supersonic collisional model (red line in top panels) the
correlation length along the perpendicular direction is much shorter than along
the parallel one, indicating strong anisotropy of the RM structures.  In all
remaining collisional models the correlation lengths along both directions are
comparable, demonstrating very small degree of anisotropy of RM. We can
recognize, that for a given sound speed, the parallel correlation length
increases with the strength of mean field. This is not observed in perpendicular
direction for supersonic models, for which the correlation length are almost
insensitive to $B_{ext}$ (red and blue lines in top right panel).

For collisionless models a different trend is observed (see lower panel in
Fig.~\ref{f6af1}). Model~5 (red line), which best resembles ICM, shows strong
anisotropy with parallel correlation length $l_\parallel \sim 39$ and
perpendicular one $l_\perp \sim 22$, on the contrary to the corresponding
collisional model, which is almost isotropic (see green line in the upper
panels). The parallel correlation for this model is the largest among our
collisionless models. The shortest correlation, both in the parallel and
perpendicular directions, is observed in model~4 (blue line), which is
supersonic super-Alfv\'enic. Still, the perpendicular correlation length is
almost twice as large as the parallel one for this case. Generally, for models
with small $c_\parallel$ (models 3, 4, and 6), the correlation lengths in both
directions are shorter, although they preserve some degree of anisotropy, and
they depend on mostly on the value of $c_\perp$. For models 1 and 2
($c_\parallel = 1.0$), the correlation length in both directions scales with
$c_\perp$, as well, except for model~5, which has the largest parallel, but the
shortest perpendicular correlation length among this group of models. These
observation indicate that the instabilities prompted in the systems lead to more
complex and smaller size configurations of the magnetic field.

\section{Discussion}
\label{sec:discussion}

\subsection{Collisionless MHD turbulence}
\label{sec:discussion:turbulence}

Concerning the turbulence cascade, two main differences can be pointed out
between the collisionless and collisional MHD models.

The first is due to the differences in the phase speed of the wave modes
introduced by the anisotropy in pressure. In fact, the linear Alfv\'en wave
velocity may be strongly modified by the pressure anisotropy, and eventually
become unstable (see Eq.~\ref{eq:lin_waves}). The concept of
super/sub-Alfv\'enic turbulence, for instance, can now be misleading in the
sense that it does not reflect anymore the dynamical importance of the magnetic
field. These changes are only important in the high $\beta$ plasma regime,
though. In the low $\beta$ regime, the role of the thermal pressure (and
consequently of its anisotropy) is secondary in the turbulence dynamics.

Second, the firehose and mirror instabilities (when present) amplify the
turbulent power at the small scales. This occurs in the unstable regime, in
which the free energy in the pressure anisotropy (at large scales) is
transformed into kinetic and magnetic fluctuations, which grow faster at small
scales. The smaller the scale the more effective this energy injection is, since
the instabilities have growth rates which are inversely proportional to the
scale. In an evolved, saturated state, the smaller the scale the larger the
local ``background'' magnetic field intensity, which quenches the instabilities.
As observed in \citet{Kowal_etal:2011}, in order to the unstable modes to
efficiently inject energy, they need to have growth times shorter than the
cascading time, otherwise they are destroyed at the beginning of their
development.

The interchange between thermal and mechanical energy is much more complex in
the collisionless case, as non-local (in scales) energy transfer is supposed to
take place. If we consider a turbulent system where the only source of energy is
the mechanical energy injected at the scale $L$, the turbulent motions create
anisotropies in pressure, depositing there some energy. Part of this energy is
again released in mechanical form by the instabilities, mainly at the
small-scales, which reduces locally the pressure anisotropy. Therefore, there is
transfer of mechanical energy from the large to the small scales, which is
obviously non-local. There are many evidences showing that the kinetic
instabilities in the microphysical scales efficiently drain the free energy from
the temperature/pressure anisotropy (caused by the large scale fluid motions),
generating entropy and quickly reducing the temperature anisotropy via anomalous
collisions over all the volume \citep[see][and references
there]{SantosLima_etal:2014}. This issue, however, is still a matter of debate
\citep[see e.g.][]{MogaveroSchekochihin:2014} which is beyond the scope of this
work and will be discussed elsewhere.

Finally, a few considerations about dissipative effects are in order. If we
completely ignore the anomalous collisions mentioned above, the physical picture
of a weakly collisional plasma implies large viscosity parallel to the field
lines for motions parallel to the field lines. At the same time, the other
viscous components and the electric resistivity are small.\footnote{This comes
from the fact that the parallel viscosity component is proportional to the ions
mean-free-path, while the resistivity is proportional to the inverse of the
mean-free-path for the electron-ion collisions.} This makes the resistive scales
far below the viscous scales for the compressible motions. Based on the
collisional estimates for these transport coefficients, the ICM is expected to
have nearly small Reynolds numbers ($Re \sim 10-100$) but very large magnetic
Reynolds numbers \citep[$~10^{20}$,][]{KunzLesur:2013}, which reflects a short inertial
range for the compressible motions. In this work we skipped these complications,
omitting the diffusive terms in the MHD equations. However, numerical
simulations always bring about effective viscosity and resistivity to the
system, which should be kept in mind. While it is difficulty to determine with
precision the dissipative range in our simulations, we can roughly estimate it
as given by scales $\lesssim 16$ cells (which for our resolution of $512^3$ is
equivalent to $k \gtrsim 32$), which is in good agreement with the velocity
power spectrum of the collisional MHD turbulent models presented in
\citet{Kowal_etal:2011} for comparison with the collisionless models).

\subsection{The applicability of a double-isothermal closure}
\label{sec:discussion:applicability}

The double-isothermal closure employed in this work presumes an infinite
reservoir of thermal free energy ($p_\parallel - p_\perp$) to be converted, by
the small-scale instabilities, into mechanical and magnetic energies. This is a
numerical (theoretical) approximation, with a conceptual similarity to that of
the mechanical driving of turbulence. Turbulence naturally decays, but the
modeling presented, in general, includes a source term that sustains the
driving at large scale. Here, the instabilities naturally evolve towards a
quasi-stable regime, but the numerical implementation of the double-isothermal
closure corresponds, in other words, to a continuous driving of pressure
anisotropy.

There is a number of physical mechanisms known to be responsible for increasing
pressure anisotropy in plasmas, such as the magnetic moment conservation in
expanding plasmas \citep{Matteini_etal:2012, Falceta-GoncalvesKowal:2015},
beaming of particles in reconnecting magnetic fields \citep{Gosling_etal:2005},
velocity drifts between ions and $\alpha$-particles \citep{Matteini_etal:2015},
and others. It is very likely that at least one, and possibly more, of these
mechanisms operates at large scales in the ICM. The consequence is a constant
driving of pressure anisotropy, which evolves as small-scale instabilities grow.
Even though the growth rate, and the saturation of instabilities to quasi-stable
regimes, is known to be fast ($\tau^{-1}_{\rm sat}\sim k r_{\rm L,i} \Omega_i$)
the presence of a continuous driving source will inevitably resulting in a
persistent marginally unstable regime for the plasma. The level of the pressure
anisotropy will naturally depend on the anisotropy driving mechanisms and their
rates. In this work this value has been conjectured, as a theoretical exercise,
however the conclusions remain given that the instabilities will operate at the
same scales independent on the pressure anisotropy levels.

\subsection{Faraday Rotation maps as diagnostic tools}
\label{sec:discussion:diagnostics}

The results of this work show that the presence of temperature/pressure
anisotropies in the plasma is able to cause the development of instabilities
that introduce changes in the statistics of the Faraday Rotation maps with
respect to the standard collisional MHD. These changes are more evident in the
small spatial scales (see Figs.~\ref{f4rm1}--\ref{f4rm4} and the power spectra
in Fig.~\ref{f2s2}). Our approach can be viewed as complementary to other works,
studying e.g. the $n-B$ correlation and mapping of RM as a function of radius
\cite[see][]{Bonafede_etal:2011, Bonafede_etal:2015}. Our numerical simulations
provide self-consistent evolution of magnetic field and density, therefore,
allow us to study the PDFs and power spectra of RM in a self-consistent way. We
do not study the radial dependence of $B$ or $\rho$, however. As we explained in
Section~\ref{sec:setup:conditions} our models represent a small box embedded in
an ICM or, in other words, the small scale part of turbulent cascade ending in
the dissipation range. This, however, rises the question if these small scale
features in FR maps could be observed with the current instruments, since they
could be smeared by the given beam resolution \citep{EnsslinVogt:2003}. The mean
free path is in range 0.05--30~kpc for Hydra A cluster
\cite{SchekochihinCowley:2006}, and 3--7~kpc in Coma cluster
\cite[see][]{Andrade-Santos_etal:2013, Sanders_etal:2013}. Therefore, we must
consider an observational spatial resolution as fine as the order of 1~kpc.
Future Square Kilometre Array should be able to reach scales of tens of pc for
the nearest clusters. Moreover, we must stress that collisionless plasma
instabilities also result in modifications in the one-point statistics of
Faraday Rotation (see Fig.~\ref{f3h2} and Table \ref{tab:vskrm}), which means
that a collection of a large number of measurements of distant background radio
galaxies would be enough to probe this effect in a nearby cluster.

The goal of our work is to determine if the FR maps obtained from observations
could manifest any sign of the collisionless plasma features, allowing us to
restrict the conditions of the ICM plasma, not only the sonic and Alfv\'enic
regimes, but also the degree of pressure anisotropy. Moreover, the FR maps
provide us, under assumption of the $B-\rho$ relation, with the estimations of
the strength of the magnetic field component parallel to LoS.
Our studies use the exact
$B-\rho$ relation, therefore allow us to test the validity of the assumptions
used to determine $B$ in ICM.

\citet{Wu_etal:2015} obtained an empirical law relating the magnetic field
intensity in the LoS with the sonic Mach number of the turbulence and the
variance of the RM measurements, which are both observable quantities. They
built synthetic FR maps employing simulated collisional MHD turbulence models
with different regimes of the plasma $\beta$ ratio and the sonic Mach number,
for the case of solenoidal turbulence forcing, and considering different angles
of view. They showed that their empirical relation could recover reasonably well
the magnetic intensity for most of the LoS. In their analysis, this empirical
relation may be understood in terms of the spectral distribution and the
correlation of the magnetic and density fluctuations, for individual MHD modes.
In the framework of the double-isothermal collisionless MHD model used in the
present study (focusing on high $\beta$ turbulence aiming at applications to the
ICM), the extraction of a similar empirical law would be impossible without a
previous knowledge of the real anisotropy level, because this may modify
drastically the MHD modes (see Figs.~\ref{fig:dispersion}
and~\ref{fig:correlation} in the Appendix) as well as the turbulence cascade
(see Section V.B).

A study of the statistics of the synthetic FR maps of collisionless models
considering different closures is desired. However, such models require a
sub-grid approach to constrain the anisotropy level due to micro-scale
processes.  As remarked before, \citet{SantosLima_etal:2014} employed a sub-grid
model parameterizing the isotropization rate, in order to study the turbulent
statistics and dynamo in the ICM. This may have also possible impact on the
determination of the transport properties of the plasma
\citep{SchekochihinCowley:2006, SantosLima_etal:2014, Kunz_etal:2012} and cosmic
rays diffusion (through their interactions with the turbulence;
\cite[see][]{YanLazarian:2002}. The anisotropies and plasma instabilities may
also have an important effect on particle acceleration by turbulence in galaxy
clusters \cite[see][]{BrunettiLazarian:2011a, BrunettiLazarian:2011b,
Miniati:2015}. The properties of temperature isotropization in collisionless
plasmas are currently being explored in extensive theoretical and numerical
studies \cite[e.g.,][]{Schekochihin_etal:2012, Riquelme_etal:2015,
Kunz_etal:2014, SantosLima_etal:2014}. For the case of dynamo action,
\cite{BhatSubramanian:2013} have simulated a fluctuating dynamo with varying
magnetic Reynolds numbers finding that the intermittent magnetic field
contributions to the FR are still significant. Following the authors, the strong
field regions contribute to only $15-20$\% of the RM, showing that the main
contribution comes from the fluctuating magnetic fields. We intend to explore
this issue elsewhere.

The precise characteristics of the magnetic fields in galaxy clusters is still
unknown. As we already mentioned in the beginning of this subsection, the
forthcoming large radio telescopes will open a new era in the observation of
these fields and should help to understand their origin and structure. The new
Low Frequency Array (LOFAR) and the planned Square Kilometre Array (SKA) trace
low-energy cosmic ray electrons allowing us to map the structure of weak
magnetic fields in many regions through FR measures \citep{Beck:2015}. In this
sense,  the use of FR maps as diagnostic tool could lead to a better
understanding and interpretation of the observations. \citet{Bonafede_etal:2015}
have studied the capabilities of the SKA in constraining the properties of
magnetic fields inside and around galaxy clusters. They showed that this
instrument will be able to recover scales in the magnetic field properties of
the ICM much smaller than the present instruments, which would be finally
comparable to the simulated data available.

\section{Summary and conclusions}
\label{sec:conclusions}

We have performed numerical simulations within two different formalisms, namely
the standard collisional MHD, and a double-isothermal collisionless MHD model
that incorporates the effects of pressure anisotropy which lead to the
appearance of the firehose and mirror instabilities in the dynamics of the
plasma. We have analyzed the magnetic field and the Faraday Rotation maps along
different lines of sight, and carried out an extensive statistical study of
these two quantities including the power spectrum, the probability distribution
function, and two-point correlation functions. In order to better understand the
dependence of these quantities on the turbulence regime in which the plasma
evolves, we have performed the simulations for six different models covering the
sub/super-Alfv\'enic and sub/supersonic regimes, including a model that matches
the conditions prevailing in compressed regions of the intracluster medium of
galaxies.

Our results show that important imprints of the pressure anisotropy may be
present in the magnetic fields and their associated Faraday Rotation maps. In
particular, we find that the magnetic field in the collisionless MHD approach
may show a more granulated structure than its MHD counterpart and its spectrum
may present an excess of power at the small scales due to the enhancement of
magnetic fluctuations originated from the firehose instability (which
corresponds to physical scales of few to tens of kiloparsecs). This is also
evidenced in the Faraday Rotation maps and in the correlation lengths extracted
from two-point functions, which turn out to be smaller in collisionless MHD.
This feature is particularly evident in the model corresponding to the
conditions prevailing in compressed regions of the intracluster medium (see
Model 5 in related figures).

These imprints should be visible at small scales of the flow and may be below
the limit of detectability of the current observational tools. As mentioned in
Section \ref{sec:discussion:diagnostics}, the SKA telescope will be able to
probe such small scale fluctuations in the ICM \citep{Bonafede_etal:2015}.

Finally, we should remark that the results above were obtained by neglecting the
plasma feedback on the instabilities. Plasma particle scattering by the
electromagnetic fluctuations of the instabilities may cause their saturation and
the relaxation of the pressure anisotropies \citep{SantosLima_etal:2014}. If
included in the numerical simulations, this relaxation may wash out the
instabilities thus further constraining their impact on the magnetic field
distribution or the Faraday Rotation maps. We will explore this issue in a
forthcoming work.

\section*{Acknowledgements}
MSN acknowledges support from a grant of the Brazilian Agency FAPESP
(2010/50298-8) and from a grant of the Argentinian National Council of
Scientific and Technic Investigation CONICET. GK acknowledges support from
FAPESP (grants no. 2009/50053-8, 2013/04073-2 and 2013/18815-0) and CAPES (PNPD
1475088). EMGDP ackowledges support from FAPESP (2013/10559-5, and 2011/53275-4)
and CNPq (306598/2009-4) grants RSL ackowledges support from FAPESP
(2013/15115-8, and DFG thanks the European Research Council (ADG-2011 ECOGAL),
and Brazilian agencies CNPq (no. 302949/2014-3), CAPES (3400-13-1) and FAPESP
(no.2013/10559-5) for financial support.




\bibliographystyle{mnras}
\bibliography{manuscript} 




\appendix

\section{Firehose and Mirror Instabilities}
\label{appendix:instabilities}

\renewcommand{\thefigure}{A\arabic{figure}}

The linear analysis of the double-isothermal collisionless MHD equations (see
eqs.~\ref{eq:mass} to~\ref{eq:lin_waves}) reveals that they allow for the
occurrence of the firehose and mirror instabilities which are described below.

\subsubsection{Firehose instability ($c_{\parallel} > c_{\perp}$)}

Defining:
\begin{equation}
f = c^2_{\perp}/c^2_{\parallel} - 1 + V^2_{A}/c^2_{\parallel}.
\end{equation}

When $f < 0$, the Alfv\'en modes become unstable and there is no wave propagation.
This happens because the tension force resisting to the bending of the field
lines disappears The growth rate of the firehose instability associated to this
unstable Alfv\'en modes is given by:
\begin{equation}
\left(\gamma_{f,A}\right)^2 = c^2_{\parallel} \lvert f \rvert k^2 \cos^2 \theta,
\end{equation}
showing that the fastest growing mode is parallel to the background magnetic
field ($\theta = 0$).

For specific angles of propagation, the slow modes also become unstable for $f <
0$. These angles are in the interval:
\begin{equation}
0 < \theta < \arccos \sqrt{ 1 + f / \left(1 - c^4_{\perp}/c^4_{\parallel} \right) },
\end{equation}
and the maximum growth rate will be for the mode parallel to the background
magnetic field (i.e., $\theta = 0$, which is in the limit of incompressible
pseudo-Alfv\'en modes):
\begin{equation}
\theta_{\max} = 0, \;\;\;\;\;
\left(\gamma_{f,S} \right)^2_{\max} = c^2_{\parallel} \lvert f \rvert k^2.
\end{equation}

\subsubsection{Mirror instability ($c_{\perp} > c_{\parallel}$)}

We define:
\begin{equation}
m = - c^2_{\perp}/c^2_{\parallel} + 1 + V^2_{A}/c^2_{\perp}.
\end{equation}

In the regime $m < 0$ (which implies $c_{\perp} > c_{\parallel}$), the slow
waves can become unstable for some propagating angles, giving rise to the mirror
modes. The unstable modes have angles in the interval
\begin{equation}
\arccos \sqrt{ \max \left[ 1, - \frac{1}{2} m \left( c^2_{\parallel} / c^2_{\perp} \right) / \left(1 - c^4_{\parallel}/c^4_{\perp} \right) \right]} < \theta < \pi/2.
\end{equation}

The maximum growth rate will be given by:
\begin{itemize}
\item $0 > m > - 2 \left( 1 - c^4_{\parallel}/c^4_{\perp} \right) \left( c^2_{\perp}/c^2_{\parallel} \right) $
\begin{equation}\nonumber
  \theta_{\max} = \arccos \sqrt{ - \frac{1}{2} m \left( c^2_{\parallel} / c^2_{\perp} \right) / \left( 1 - c^4_{\parallel} / c^4_{\perp} \right) },
\end{equation}
\begin{equation}\nonumber
\left( \gamma_{m,S} \right)^2_{\max} = \frac{k^2}{2}
\left\{
\sqrt{ \left( c^2_{\perp} + V^2_{A} \right)^2
+ c^4_{\parallel} m^2 / \left( 1 - c^4_{\parallel} / c^4_{\perp} \right)}\right\}-
\end{equation}
\begin{equation}
- \left( c^2_{\perp} + V^2_{A} \right) \frac{k^2}{2}
\end{equation}

\item $m < - 2 \left( 1 - c^4_{\parallel}/c^4_{\perp} \right) \left( c^2_{\perp}/c^2_{\parallel} \right)$
\begin{equation}\nonumber
\theta_{\max} = 0,
\end{equation}
\begin{equation}
\left( \gamma_{m,S} \right)^2_{\max} =
\frac{1}{2}
\left\{ \left| 2 c^2_{\parallel} - c^2_{\perp} - V^2_{A} \right|
- (c^2_{\perp} + V^2_{A}) \right\} k^2.
\end{equation}

\end{itemize}

The mirror modes conserve the conventional property of the slow modes, i.e.,
they allow for a negative correlation between the density fluctuations and the
magnetic field component parallel to the background magnetic field.

\section{Linear dispersion for the initial conditions of the Models}
\label{appendix:dispersion}

\renewcommand{\thefigure}{B\arabic{figure}}

\begin{figure*}
 \centering
 \includegraphics[width=0.9\textwidth]{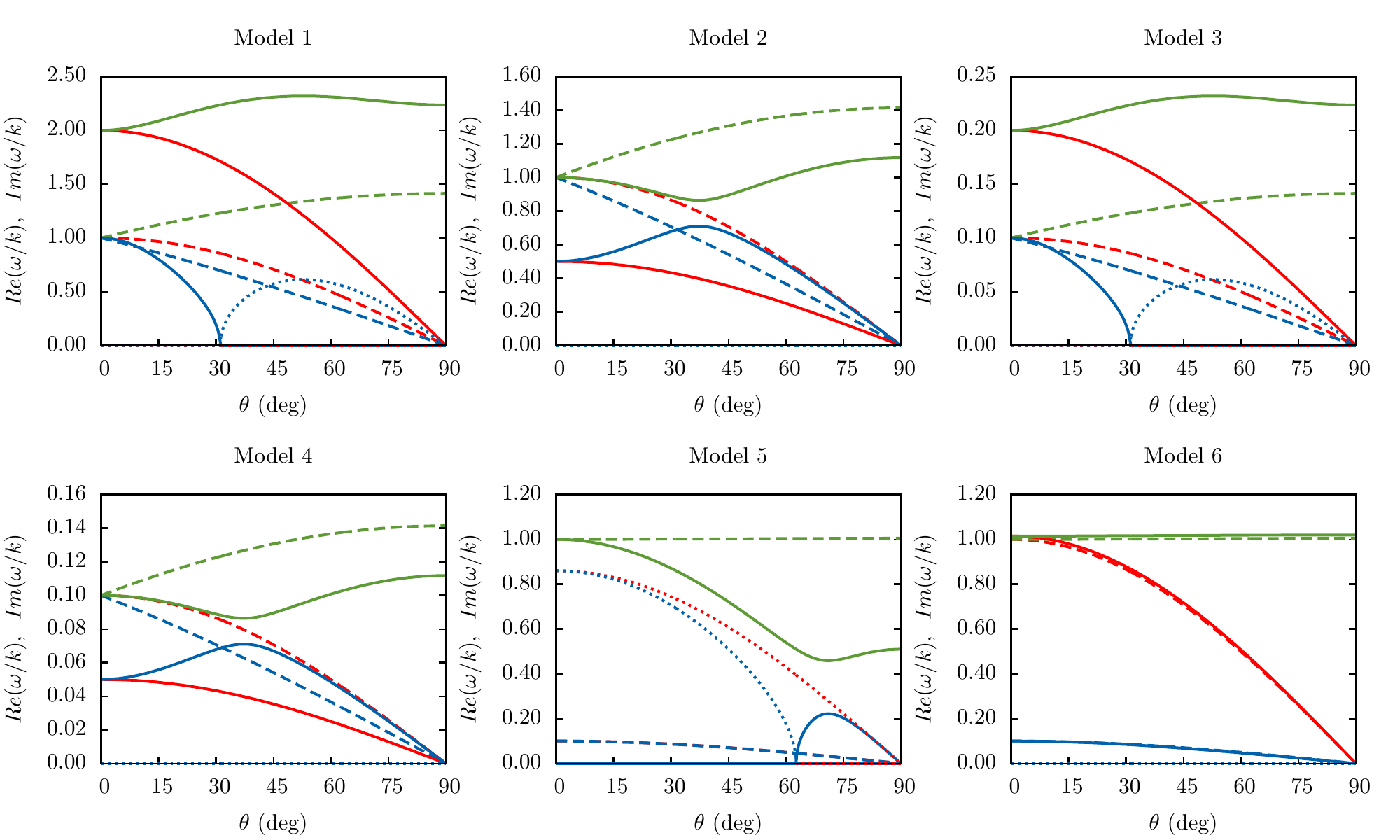}
 \caption{Real (continuous lines) and imaginary (dotted lines) linear phase
velocities for different waves considering the initial conditions of the
simulated models (see Table~\ref{tab:models}). The phase speeds for the comparative
MHD models are given by dashed lines. Each color represents a different wave:
Alfv\'en (red), fast (green), and slow magneto-sonic (blue).}
 \label{fig:dispersion}
\end{figure*}

\begin{figure*}
 \centering
 \includegraphics[width=0.9\textwidth]{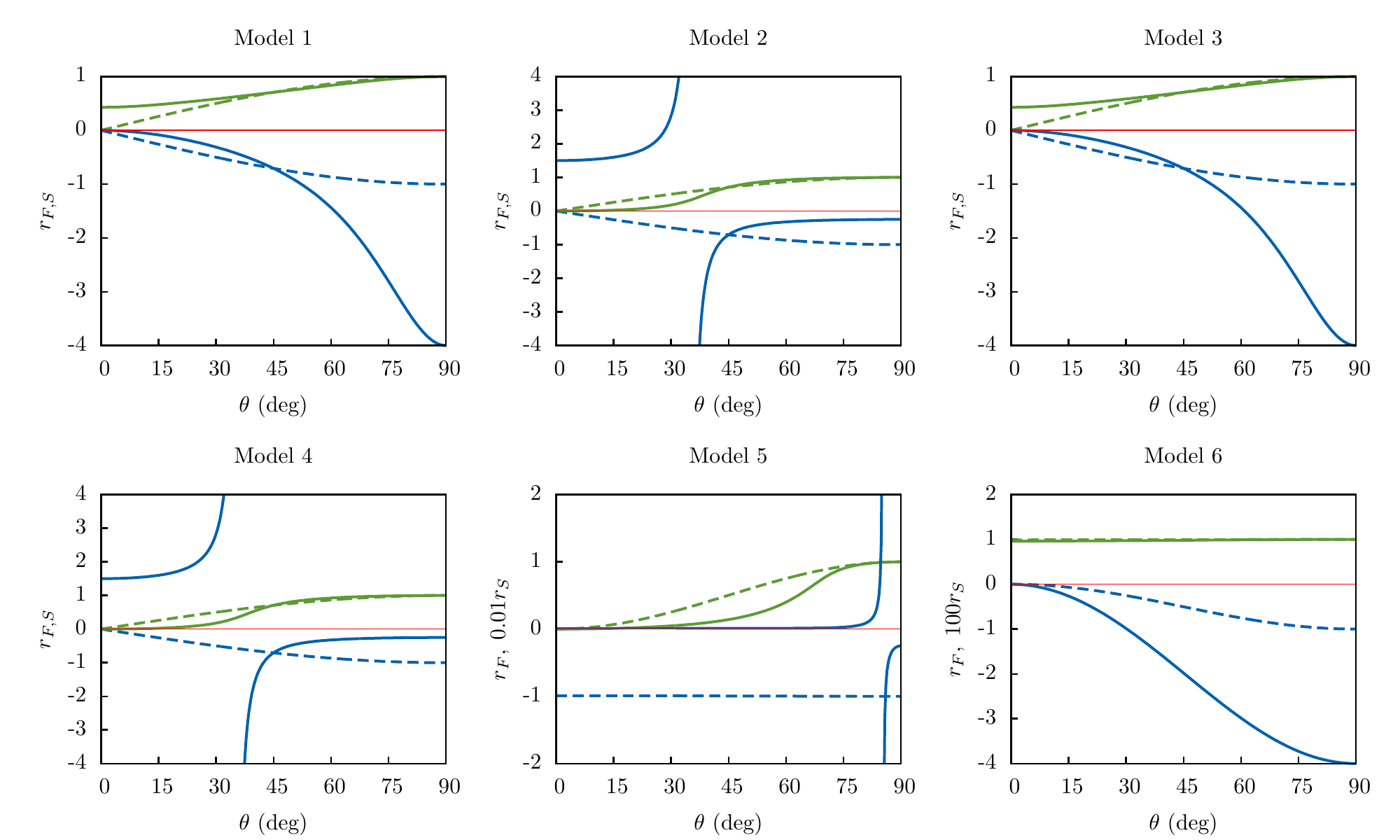}
 \caption{The lines show the magnetic-density correlation
(eq.~\ref{eq:correlation}) for the linear modes slow (blue) and fast (green) as
a function of the propagation angle, calculated for the initial conditions of
the simulated models. The solid lines represent the anisotropic pressure
collisionless models and the dashed lines represent the comparative collisional
MHD models (see Table~\ref{tab:models}).}
 \label{fig:correlation}
\end{figure*}

Figure~\ref{fig:dispersion} shows the real and imaginary phase velocities of the
linear waves as a function of the propagation angle (see dispersion relation in
equation~\ref{eq:lin_waves}) for the initial conditions of each of the simulated
models in this work (see Table~\ref{tab:models}).

Figure~\ref{fig:correlation} depicts the density-magnetic field fluctuation
correlation for the compressible linear modes as a function of the propagation
angle, also for the initial conditions of each model of Table~\ref{tab:models}. This
correlation was calculated using equation 9 of \citet{HauWang:2007}:
\begin{equation}\label{eq:correlation}
r_{F,S} \equiv \left( \frac{\delta B}{\delta \rho} \right)_{F,S} \left( \frac{B_{ext}}{\rho_0} \right) =
\frac{a \pm \sqrt{\Delta}}{2(a + c^2_{\perp})},
\end{equation}
where $a = V^2_A - c^2_{\perp} + 2 \left( c^2_{\perp} - c^2_{\parallel} \right)
\cos^2 \theta$, the $\pm$ are for the fast ($F$) and slow ($S$) modes, and
$\Delta$ is given by eq.~\ref{eq:delta}.


\bsp	
\label{lastpage}
\end{document}